\documentclass[final]{msml2021} 

\def\E#1#2{\mathbb{E}_{#1}\left[#2\right]}

\def\IC{\mathbb{C}}

\def\IZ{\mathbb{Z}}
\def\IR{\mathbb{R}}
\def\IP{\mathbb{P}}

\def\CM {{\cal M}}
\def\CN {{\cal N}}

\def\CP {{\cal P }}

\def\CL {{\cal L}}

\def\CO {{\cal O}}
\def\CX {{\cal X}}

\def\CE {{\cal E}}
\def\CG {{\cal G}}

\def\CK{{\cal K}}

\def\CY{{\cal Y}}

\newcommand{\eq}[1]{Eq.~(\ref{eq:#1})}

\newcommand{\be}{\begin{equation}}
\newcommand{\ee}{\end{equation}}
\newcommand{\ba}{\begin{array}}
\newcommand{\ea}{\end{array}}
\newcommand{\bea}{\begin{eqnarray}}
\newcommand{\eea}{\end{eqnarray}}

\renewcommand{\Im}{{\rm Im }}
\renewcommand{\Re}{{\rm Re }}

\def\one{{\hbox{ 1\kern-.8mm l}}}

\def\p{\partial}
\def\ba{\bar{a}}

\def\bz{\bar{z}}
\def\bZ{\bar{Z}}

\def\bj{{\bar{j}}}
\def\bi{{\bar{i}}}

\def\bs{{\bar{s}}}

\def\bx{\bar{x}}

\def\bz{\bar{z}}

\def\bJ{{\bar{J}}}

\def\bZ{\bar{Z}}

\makeatletter
\let\Ginclude@graphics\@org@Ginclude@graphics
\makeatother
\title[Numerical Calabi-Yau metrics from holomorphic networks]{Numerical Calabi-Yau metrics from holomorphic networks}
\usepackage{times}

\begin{document}
\msmlauthor{%
 \Name{Michael R. Douglas$^*$} \Email{mdouglas@scgp.stonybrook.edu}\\
 \Name{Subramanian Lakshminarasimhan$^\dag$} \Email{subramanian.lakshminarasimhan@stonybrook.edu}\\
 \Name{Yidi Qi} \Email{yidi.qi@stonybrook.edu}\\
 \addr Department of Physics, YITP and SCGP, Stony Brook University\\
 {$\dag$ \addr Department of Applied Mathematics and Statistics, Stony Brook University}\\
{$*$ \addr Center of Mathematical Sciences and Applications, Harvard University}
}

%
\maketitle

\def\Large{\large}
\def\LARGE{\large\bf}


\begin{abstract}
We propose machine learning inspired methods for computing 
numerical Calabi-Yau (Ricci flat K\"ahler)
metrics, and implement them using Tensorflow/Keras.
We compare them with previous work, and find that
they are far more accurate for manifolds with little or no symmetry.
We also discuss issues such as overparameterization
and choice of optimization methods.

\end{abstract}

\section{ Introduction }

K\"ahler manifolds with Ricci flat metrics, known as Calabi-Yau manifolds, are the
first and still the most important starting point for compactification of string theory to produce
realistic physical models.  Their study led to the discovery of mirror symmetry and a great deal
of interesting mathematics.  A brief survey appears in \citet{Douglas2015}, and textbooks
on mirror symmetry include \citet{MS1,MS2}.

No closed form expressions are known for these Ricci flat metrics, and it is generally believed that
none exist.\footnote{
But see \citet{Gaiotto2010,Kachru2020} for an analytic approach to the K3 metric.}
One can get an approximate description using methods of numerical general
relativity, as pioneered in \citet{Headrick2005}.  The special properties 
of K\"ahler geometry lead to many simplifications, starting with the representation of the metric by
a single function.  Donaldson introduced many ideas such as a representation by projective embeddings
and approximation by balanced metrics \citep{Donaldson2009}.  Subsequent works simplified and improved
the numerical methods \citep{Douglas2007,Douglas2008,Doran2008,Bunch2008,Seyyedali2009,Braun2008,Braun2008a,Headrick2013,Anderson2010,Anderson2010a,Anderson2012}
so that a fairly simple program can get accurate results.

To simplify a bit, the approach used in these works is to expand the K\"ahler potential in a polynomial basis.
As is the case for spectral methods, this is very good for approximating smooth functions with variations on
length scales $1/k$, where $k$ is the order of the polynomials used.  This is sometimes good enough, but by varying
moduli one can easily produce metrics with structure on arbitrarily short scales, say by approaching a singular
limit, as was studied in \citet{Cui2020}.  And since on a $D$-dimensional manifold, the number of basis functions grows as $\CO(k^D)$, one cannot take $k$ very large (the ``curse of dimensionality'').  Indeed, almost all of these works restrict
attention to manifolds with large discrete symmetry groups to get a manageable number of coefficients.  An exception is \citet{Braun2008} which studied
a randomly chosen quintic with no symmetry.

In recent years, there has been much success with machine learning (ML) inspired approaches to scientific
computation, and especially the solution of PDE's such as Navier-Stokes and the Schr\"odinger equation.
This is a rapidly developing area and any reviews we cite here would
become dated very quickly, but to give
a simple illustration of the approach
we follow we cite \citet{Hoyer2019}.  This work solves problems in structural optimization -- mathematically, the
unknown is a single function in a bounded domain of $\IR^2$ and the solution is the minimum of a nonlocal
energy functional (the integrated stress).  What is novel is to represent the unknown function,
not in terms of finite elements or a Fourier basis, but as the output of a neural network.  This provides
a space of approximating functions parameterized by the network weights, which can be efficiently computed
using technologies developed for machine learning.  Standard numerical optimization applied
to the energy considered as a function of the weights leads to high quality solutions, which are not limited
by the constraints of previous methods.

These methods have only begun to be explored for higher dimensional geometry.
The idea we propose and study here is to 
represent the K\"ahler potential in terms of the output of a {\bf holomorphic network} or a {\bf bihomogeneous network}.
Both are
feedforward neural networks, whose inputs are simple
functions of the complex coordinates,
which use the activation function $z\rightarrow z^2$,
and whose outputs are efficiently computable functions of the weights.
We then find the metric in this class which is closest to Ricci flat
by numerical optimization of an error function with respect to the weights, 
much as was done for highly symmetric metrics in \citet{Headrick2013}.

For the holomorphic network, the inputs are complex coordinates or low
degree sections,
the weights and intermediate values (or ``activations'') are complex, and the
 outputs are a subspace of the space of holomorphic sections.  One then
 takes a hermitian 
combination of these outputs and their complex conjugates 
to get a K\"ahler potential.

By contrast, in the bihomogeneous network, the inputs 
are the real and imaginary parts of products of holomorphic and
antiholomorphic functions, homogeneous in each separately.  Thus the
weights and intermediate values can be taken to be real.
The output of
the network is then used as a K\"ahler potential.  This turns out to work
better than the holomorphic network for reasons we will explain.

Since each layer doubles the degree of the section, 
in principle one can work with large values of $k$.
One faces the standard challenges of deep networks such as exploding gradients, 
but we were able to use depths up to 5 without this being a problem.  

Our formulation of the numerical Calabi-Yau metric problem 
is formally very similar to supervised learning -- both problems amount to fitting a function given
a set of input-output values $(x_i,y_i)$, where $x_i$ is sampled from an input distribution and $y_i$ is known.
Thus most of the task is already implemented in standard ML packages, and one only needs a few domain dependent additions.
Our code uses TensorFlow, and is available on Github.\footnote{https://github.com/yidiq7/MLGeometry}

We obtain results for a variety of quintic hypersurfaces, including the familiar
one-parameter Dwork family, and also hypersurfaces with little or no symmetry.
Our main focus is to understand the accuracy of these metrics and its dependence on parameters,
both those of the manifold and
hyperparameters (parameters of the model and learning algorithm).
We can get mean errors (MAPE defined below) around $10^{-3}$ for metrics
with no symmetry, about 100 times better than \citet{Braun2008}.
We also discuss the nature of the optimization landscape and
the overparameterized regime, which have been
important themes in recent studies of ML.

\subsection{ Related work }

Recent works which study numerical Calabi-Yau geometry include \citet{Ashmore2020,ashmore_eigenvalues_2020,anderson_moduli-dependent_2020}.
In \citet{Ashmore2020}, the K\"ahler potential is represented by a gradient boosted decision tree.
The problem is solved at various lower degrees $k$, and the ML model does extrapolation in the degree
(Richardson extrapolation).    
In \citet{ashmore_eigenvalues_2020} eigenvalues of Hodge Laplacians
are computed.
The work \citet{anderson_moduli-dependent_2020} (appearing simultaneously with ours)
studies metrics with $SU(3)$ structure using neural networks.

Polynomial activation functions have been studied in many ML works,
for example \citet{Chrysos2020,Mannelli2020},
but our motivations (contact with higher dimensional complex geometry) are new.

A brief mathematical summary of the approach will appear in \citet{Douglas2020}.

\section{ Numerical complex geometry using machine learning methods}

In this section we alternate between the complex geometry needed for our problem and review of the concepts
we use from machine learning.  For the benefit of readers who are not experts in both fields, we review some basic concepts on both sides in \appendixref{appendix:reviews}.
Some useful background material includes \citet{griffiths2014principles,huybrechts2005complex} on
complex geometry and \citet{bishop2013pattern} on machine learning.  

\subsection{ Brief review of the Calabi-Yau metric problem }

Our problem is the following: we are given a complex manifold $M$
with local coordinates $z^i\equiv (z^1,\ldots,z^n)$ and their complex
conjugates $\bz^\bi\equiv (z^1,\ldots,z^n)^*$.
which admits K\"ahler metrics, determined (as we review in \appendixref{appendix:kahler})
by a K\"ahler potential $K$ as
\be \label{eq:defKahler}
g_{i\bar j} = \frac{\partial^2\, K}{\partial z^i \partial {\bar z}^{\bar j}}
\equiv \partial_i\bar\partial_\bj K.
\ee
We want to find a K\"ahler metric which to a good approximation satisfies
Einstein's equation
\be\label{eq:einstein}
\mbox{Ricci}_{i\bar j} = \Lambda\, g_{i\bar j} ,
\ee
where $\Lambda$ is a real constant.
General arguments determine the sign of $\Lambda$ in 
terms of the topology of $M$ (its first Chern class).
Deep mathematical theorems of Yau, Aubin,
Donaldson-Song, and others, show that a solution always exists for $\Lambda\le 0$,
and give the necessary and sufficient conditions for $\Lambda>0$.
For $\Lambda\ne 0$ the solution is generally unique, while for $\Lambda=0$
it is unique up to an overall rescaling $g_{i\bar j}\rightarrow R^2 g_{i\bar j}$
for $R\in \IR^+$.  The $\Lambda=0$ metrics are called ``Ricci flat.''  While we focus
on this case, the methods we discuss can be used for the other cases.

In this work we take $M$ to be the quintic hypersurface in $\IC\IP^4$, 
the set of solutions to a polynomial equation such as
\be \label{eq:fermat}
0 = f(Z^1,Z^2,Z^3,Z^4,Z^5) = \sum_{i=1}^5 (Z^i)^5 + \psi Z^1 Z^2 Z^3 Z^4 Z^5
\ee
where $\psi$ is a fixed complex number (a ``complex modulus'').
Here $Z^i$ are projective coordinates, and a point on $\IC\IP^4$ is an equivalence
class of points in $\IC^5-\{\vec 0\}$ under 
$Z^i\sim\lambda Z^i\,\forall\,\lambda\in\IC$.
$M$ is a three complex dimensional space which is a manifold,
except at possible singular points $Z$
at which
\be \label{eq:disc}
0 = f(Z) = \partial_i f(Z) \,\,\forall 1\le i\le 5.
\ee
For \equationref{eq:fermat},
these only exist for $\psi=-5\omega$ where $\omega^5=1$ is a fifth root of unity.
For other $\psi$, one can show that $c_1(M)=0$, so $\Lambda=0$ in \equationref{eq:einstein}.
This Fermat quintic (for $\psi=0$) or Dwork family of quintics
is the most studied example, {\it e.g.} see \citet{Candelas1991}.

More generally, one can take $f$ to be any quintic polynomial.  Such a polynomial
has 126 adjustable coefficients, and two manifolds $M_1$ and $M_2$ defined in
terms of two polynomials $f_1,f_2$ are equivalent (complex diffeomorphic) only
if there exists a linear transformation $W^i\equiv L^i_j Z^j$ such that
$f_1(Z)=f_2(W)$.  This introduces 25 redundancies in the parameterization, so
the space parameterizing quintic hypersurfaces in $\IC\IP^4$ 
(the complex moduli space) is 101 complex dimensional.
One can show that solutions of \equationref{eq:disc} only exist on a codimension one variety
in this space (the ``discriminant locus'' $\Delta$), so the moduli space of nonsingular
quintic CY manifolds is 101 dimensional.

Of the many generalizations of this construction, the most used in string theory
are to replace $\IC\IP^N$ by products of projective spaces, weighted projective
spaces and toric varieties.  
Here we simply point out that the constructions below can be
straightforwardly adapted, by replacing homogeneity with
weighted multihomogeneity, or equivalently gauge invariance in the GLSM
constructions.  Many details of these generalizations can be found in the
works of the Penn group \citep{Braun2008a,Anderson2010}.

\subsection{ The embedding method and geometric quantities}
\label{ss:geom}

Following \citet{Donaldson2009}, most work on numerical Calabi-Yau metrics
represents the metric using an embedding by holomorphic sections of a very ample line bundle $\CL$.  This embedding is a map into a linear space, analogous to
spectral embeddings such as the ``Laplacian eigenmap'' construction, but with the great advantage that the map has a simple exact form.
A review of the embedding method is in \appendixref{appendix:embedding}.

The embedding representation gives us
a natural family of metrics, the pullbacks of the
Fubini-Study metrics from complex projective space. 
Using our embedding by a basis of $N$ sections $s^I$, and
pulling back a Fubini-Study metric on $\IC\IP^{N-1}$, 
the embedding then leads to the K\"ahler potential
\be \label{eq:a1}
K = \log \sum_{I,\bJ} h_{I,\bJ} s^I \bs^\bj
\ee
where $s^I$ is a basis of $N=h_0(\CL)$ holomorphic sections.  
This gives us an $N^2$ real dimensional
family of metrics parameterized by the hermitian matrix $h_{I,\bJ}$.

The geometry of a Calabi-Yau manifold (let $n$ be its complex dimension;
eventually we will take $n=3$) is
determined by two fundamental differential forms.  The first is 
present on any K\"ahler manifold -- it is the K\"ahler form
\be
\label{eq:omegag}
\omega = \partial_i \partial_\bj K \, dZ^i \wedge d{\bar Z}^{\bar j} .
\ee
This carries the same information as the metric and can be used to write
the volume element $d\mu_g = \sqrt{g}$. 
On a K\"ahler manifold, one can do this without taking square roots: 
\be
\label{eq:MA}
d\mu_g \equiv \omega_g^{\dim_\IC M} =
\det \omega_g = \det_{i,\bj} \partial_i \partial_\bj K .
\ee
The other, which is only present for a Calabi-Yau manifold, is the
holomorphic $n$-form
\be
\Omega = \Omega_{i_1\ldots i_n} dZ^1 \wedge \ldots \wedge dZ^n .
\ee
It is nonvanishing and nonsingular, so we can define an associated volume form
\be \label{eq:Omegavol}
d\mu_\Omega \equiv \CN_\Omega \Omega \wedge \bar\Omega .
\ee
The normalization constant $\CN_\Omega$ will be set to make the total integrals
of both volume forms equal.
One can show that the integral $\int_M\omega_g^n$ is a topological invariant, 
which stays fixed as we continuously vary the metric.  Thus we can begin by computing the value of $\CN_\Omega$ which does this
for the initial metric, and leave it fixed as we
search for the Ricci flat metric.

The form \equationref{eq:Omegavol} depends on the complex structure but 
is independent of the K\"ahler form
and thus the embedding we use to represent $M$.  Often one can
write an explicit formula for it -- for a hypersurface it is
\be \label{eq:nform}
\Omega = \frac{ dZ^1 \wedge \ldots \wedge dZ^{n}}{ \partial f/\partial Z^{n+1} }
\ee 
where $f$ is as in \equationref{eq:fermat}.  This formula becomes singular where
$\partial f/\partial Z^{n+1}=0$.  To fix this, one can check that one gets 
the same $\Omega$ by choosing any of other the coordinates
to play the role of $Z^{n+1}$.  One can then define several coordinate patches,
each with one of these representatives, and whose union covers $M$.

Now, we come to a very helpful simplification which follows from Calabi-Yau
geometry.  For a general metric, \equationref{eq:einstein} is a system of PDEs determining the individual metric components and
with general covariance.  It is not elliptic unless one properly fixes the coordinate system, which is quite nontrivial to do.
By using complex coordinates, \equationref{eq:defKahler} for the metric, and \equationref{eq:defKahler2} for the Ricci tensor,
it becomes a PDE for a single function without these
issues.  
\be \label{eq:PDE1}
0 = \frac{\p^2}{\p Z^i\bar\p \bZ^\bj} \log \det_{i,\bj} 
\frac{\partial^2 K}{\partial Z^i \bar\partial \bar Z^j }.
\ee
However this is still nonlinear and involves many derivatives.
But for a Ricci flat K\"ahler metric, one can integrate this
formula twice to obtain
\be \label{eq:cycond} 
 d\mu_\Omega = d\mu_g
\ee
or equivalently
\be
1 = \eta \equiv \frac{ d\mu_g }{ d\mu_\Omega } 
\ee
where the constants of integration are determined by global consistency.\footnote{
A short argument for this is that
if it were not the case, then $1/\eta$ in \equationref{eq:defeta}
would define a non-constant harmonic function on $M$, but
on a compact $M$ this does not exist.
}
Another advantage of this condition is that it
can be obtained from a convex energy functional (see \S \ref{ss:genimpl}).

To make this completely explicit, use
\equationref{eq:nform}, use the constraint $f=0$ to solve for $i,j=n+1$
in the determinant in terms of the other derivatives,
and make a final rearrangement, to get 
\be
\label{eq:defeta}
1 = \eta \equiv \frac{ 1}
{ \CN_\Omega} |\partial f/\partial Z^{n+1}|^2 \,
\det_{1\le i,j\le n} L^{i'}_{i} \frac{\partial^2 K}{\partial Z^{i'}
\bar\partial \bar Z^{j'}  } (L^\dag)^{j'}_j,
\ee 
where the matrix $L$ is given explicitly in \equationref{eq:defLmatrix}.

\subsection{ Multilayer holomorphic embeddings }
\label{ss:holo}

The idea we will pursue in this work is to use the feed-forward network:
\be \label{eq:defMLP}
F_w = W^{(d)} \circ \theta \circ W^{(d-1)} \circ \ldots \circ
\theta \circ W^{(1)} \circ \theta \circ W^{(0)} ,
\ee
where the $W^{(i)}$'s are general linear transformations, and $\theta$ is the activation function,
to define a subset of the metrics \equationref{eq:a1}.  There are several ways to do this.  In this subsection
we define a network with complex weights and activations, which defines a subspace
of $H^0(\CL^k)$.  We will then restrict \equationref{eq:a1} to this subspace.
In the next subsection we will take a different approach, forming
real combinations as the inputs to the network.

Thus, here we take the input space $\CX$ to be the ambient space $H^0(\CO_M(1))$,
the space parameterized by the homogeneous coordinates $Z^i$.
Concretely, the first layer has 5 complex inputs.
We take all of the intermediate vectors 
to be complex, and we choose
$\theta(x)$ to be a nonlinear homogeneous holomorphic function.  
The simplest choice and the one we will use is to take
\be \label{eq:deftheta}
\theta(x) = x^2 .
\ee
Thus each successive layer defines a subspace of sections of degree
twice the previous layer.

To get a real valued K\"ahler potential, we replace the final layer
$W^{(d)}$ with a general linear combination of products of the sections
with their complex conjugates,
\bea \label{eq:defFK}
K(w;Z) &=&  \log\sum_{i_d=1,\bj_d=1}^{D_d} h^{(d)}_{i_d,\bj_d} Z_d^{i_d} \bZ_{d}^{\bj_d} \\
Z^{(d)} &\equiv& \theta \circ W^{(d-1)} \circ \theta \circ \ldots \circ
\theta \circ W^{(1)} \circ \theta \circ W^{(0)} Z \\
\bZ^{(d)} &\equiv& (Z^{(d)})^* .
\eea

This construction gives us a class of metrics for each choice of depth $d$ and layer widths $D_1,\ldots,D_d$,
obtained from embeddings with degree  $k=2^d$.  The total number of real weights is
\be
N_w = 2\left( D D_1 + D_1 D_2 + \ldots + D_{d-1} D_d \right) + D_d^2 .
\ee
Generally $D_i < h_0(\CL^{2^i})$ so this will not span the complete basis of sections, in other words
we have restricted the embedding and are only using a subset of metrics.  While the final layer $z_d$ is a linear
subspace of $H^0(\CL^{2^d})$, this subspace is nonlinearly parameterized by the weights.
The hope is that it is flexible enough to describe the metrics of interest.

A variation on this is to take the inputs $Z$ to be a complete basis of sections $s^I$ of degree $k_0$.  For the hypersurface $f=0$ in $\IC\IP^{D-1}$, 
the basis will be the symmetrized 
degree $k_0$ monomials, quotiented by the ideal generated by $f$ (if $k_0 \ge \deg f$).
Other combinations of layers and activation functions are of course possible, subject to the constraint
that every activation (intermediate value) is homogeneous (a section of a definite line bundle).
Thus one cannot have skip connections, but one could take other products of outputs from previous layers.

We have implemented these networks, but they turned out to have some
disadvantages compared to the approach we will discuss next.  One is
technical: ML software does not always treat holomorphic functions
as one would expect (in particular the derivative $\p/\p z$), and
one must be careful in programming a complex network.  More
importantly, one needs very wide networks to represent
simple metrics.  Consider the K\"ahler potential 
\be
K = \log \sum_i |Z^i|^2
\ee
constructed from $k=1$ sections.  This is also a point in
every space of $k>1$ metrics (up to the overall scale), as we can write
\bea \label{eq:nestk}
K &=& k\log \sum_i |Z^i|^2 = \log \left(\sum_i |Z^i|^2\right)^k \\
&=& \log \sum_{I} c_I |Z^{I_1}\cdot\ldots\cdot Z^{I_k}|^2 .
\eea
Reproducing this sum over the complete basis of sections
would require a very wide final layer.

\subsection{ Bihomogeneous embeddings }

A variation which does not have the problems we just described
is to use bihomogeneous sections, 
meaning products of holomorphic and antiholomorphic sections,
as the activations (intermediate values) of the network.  Thus, we would take
\be \label{eq:defMLP2}
 K(w;Z) = \log W^{(d)} \circ \theta \circ W^{(d-1)} \circ \ldots \circ
\theta \circ W^{(2)} \circ \theta \circ W^{(1)} (Z\bZ),
\ee
where the inputs $(Z\bZ)$ are the real and imaginary parts of the
bihomogeneous (or ``sesquilinear'') combinations $Z^I\bZ^\bJ$.
In terms of $Z^I=X^I+iY^I$ these are
\bea
&& X^I X^J + X^J X^I - Y^I Y^J - Y^J Y^I \qquad \forall 1\le I\le J \le n, \\
&& X^I Y^J - X^J Y^I \qquad \forall 1\le I < J \le n,
\eea
which in $n$ complex dimensions make up $n^2$ real components.
The output dimension is $D_d=1$.  For $d=1$ and taking $W^{(1)}$
real, this reproduces
\equationref{eq:a1} with $k=1$.

In this network, the inputs, intermediate variables 
and all of the weight matrices will be real.  
One still needs the activation function $\theta$ to be homogeneous,
and $z\rightarrow z^2$ is still the natural choice.  The total number of real weights is now
\be
N_w = D D_1 + D_1 D_2 + \ldots + D_{d-2} D_{d-1} + D_{d-1}.
\ee

Let us check that this transforms properly under a change of projective coordinates 
$z^i\rightarrow \lambda z^i$.  Then
\bea
(Z\bar Z) &\rightarrow & |\lambda|^2 \,(Z\bar Z) \\
\theta \circ W (Z\bar Z)  &\rightarrow & |\lambda|^4 \,\theta \circ W (Z\bar Z) \\
& \vdots & \\
 K(w;Z) &\rightarrow&   K(w;Z) + 2^{d-1} \log |\lambda|^2
\eea
Again, there are many variations on this construction -- any rules of combination which are
homogeneous in the bidegrees $(1,0)$ for $z$ and $(0,1)$ for $\bar z$ are allowed.

With this construction, the higher degree
space of K\"ahler potentials contains the lower degree spaces in a simple way. 
For example, \equationref{eq:nestk} can be reproduced with width 1 intermediate layers.
This construction can also represent real-valued functions as
a difference $K_1-K_2$.  Here $K_2$ could be a fixed reference K\"ahler potential
of the same degree, or the output of a second network.

A potential problem with \equationref{eq:defMLP2} is that it is not easy to enforce positive definiteness of \equationref{eq:defKahler}.
However we got good results without explicitly doing so.
Two reasons are that we do not use the inverse metric, and \equationref{eq:cycond}  forces the determinant of the metric to be positive.

\section{ Implementation }
\label{ss:genimpl}

Other than the use of a network instead of a basis of sections, our method is as used in previous works:
optimization as in \citet{Headrick2013} using
the sampling methods of \citet{Douglas2008}.  It turns out that
this is so similar to supervised learning
that we can easily adapt standard ML software to do it.\footnote{
Supervised learning was already used in computing numerical 
Ricci flat metrics by \cite{Ashmore2020}, to extrapolate the results
from Donaldson's T-map method.  We feel the approach
taken here is more straightforward.
}
Thus we review supervised learning briefly (and at more length in appendix \ref{appendix:learn}) to explain this.

\subsection{ General implementation }
\label{ss:genimp}

In supervised learning,
we have a data set of $N_{data}$ items, each of which is an input-output pair $(x_n,y_n)$.
These are supposed to be drawn from a probability distribution $\CP$ on $\CX\times\CY$.
The goal is to choose the function from $\CX$ to $\CY$
from a given set (or ``model'')
which best describes the general relation $\CP$ between input and output,
in the sense that it minimizes some definition of the expected error (an objective or ``loss'' function).
The procedure of making this choice given the data set
is called training the network.

To phrase our problem in these terms, note that in \equationref{eq:cycond},
the left hand side $d\mu_\Omega$ is a known function of the point $Z$,
while the right hand side $d\mu_g$ is an adjustable function depending on
the parameters of the metric $g$, in other words the weights.  Thus we
can take our input dataset $x_i$ to be a set of points $Z_i$, the
output we are trying to fit is the value of $d\mu_\Omega$ at these points,
call these $y_i$, and the ``model'' we are using to fit it is $d\mu_g$.
Thus the problem is exactly of the form we desired.  We are still solving
a PDE, but the derivatives are all done in the computation of $d\mu_g$.

Let us flesh out this observation.  An interpolation
problem requires a sampling procedure for the $x_i$, a model for the
$y_i$, an objective function which measures the quality of the fit, and
an optimization procedure.

As we just explained, the Ricci flatness condition is $\eta=1$.
The least squares error is then
\be \label{eq:Lquad}
\CE = \int_M d\mu_{ref} \left(\eta-1\right)^2 
\ee 
where $d\mu_{ref}$ is a ``reference measure'' on $M$.  Here $\eta$
(\equationref{eq:cycond})
is implicitly a function of the weights through $d\mu_g$; we put this
in the numerator 
to simplify this dependence.  Varying
with respect to the weights, and assuming that there are enough weights
to vary $\eta$ at every point in the support of $d\mu_{ref}$, its minimum
will be $\eta=1$ on this support. 

Note that 
the integral of any convex function $F(\eta)$ will have the same 
optimum.
Out of these choices, we would prefer to use an objective function which is convex in the parameters.  
An important feature of the continuum PDE is that if $F(\eta)$ is convex,
then considered as a functional of the K\"ahler potential, 
$\CE_F$ has a unique critical point.\footnote{
This is why this highly nonlinear
complex Monge-Ampere equation is relatively tractable. 
A short derivation is in \citet{Headrick2013},
and \citet{Berndtsson2013} has an extensive discussion of this point with references.}
However this need not be so when considered as a function of the parameters.
Indeed, it is never the case for a nontrivial
objective function and the multilayer FFN \equationref{eq:defMLP},
as is clear for the simplest example of a two layer linear
network $y=w_1w_0 x$.  

Another complicating factor is that we will use a reference
measure $d\mu_{ref}$ supported on a finite set of points.
Since the variation of $\CE_F$ only imposes constraints
on the support of $d\mu_{ref}$, the variational equations and
their solutions will generally depend on this support (but not
on the values of $d\mu_{ref}$).  Furthermore
one expects multiple or degenerate solutions in the overparameterized regime we discuss later.

In practical ML, while nonconvexity and nonuniqueness
can lead to problems, they turn out not to
be as serious as one might expect.  This point has received extensive
study and can be understood analytically for linear networks (with
activation function $f(x)=x$), see for example \citet{Advani2020}.
But in practice, one varies the initial conditions and learning rate
until one finds choices which work, as discussed in textbooks
such as \citet{Goodfellow-et-al-2016}.
This is part of the ``art'' of machine learning.

As for the objective function,
least squares as in \equationref{eq:Lquad} is usually a good choice,
though we consider alternatives in \S \ref{ss:experiments}.
As a reference measure, we take the Monte Carlo
measure (or ``empirical measure'')
defined as an average over a set of $N$ randomly chosen points on $M$,
\be
d\mu_{ref}(x) = \frac{1}{N} \sum_{i=1}^N \delta(x-x_i) .
\ee 
If $x_i$ is sampled from some $dP$, then 
$\lim_{N\rightarrow\infty} d\mu_{ref}\rightarrow dP$
under very weak conditions on $dP$ and the integrand.

To sample points on $M$ we use the procedure of \citet{Douglas2008}.
We sample pairs of points $a,b$ from a normal distribution on $\IC^5$.  Regarded as
points on $\IC\IP^4$, they are distributed according to \equationref{eq:defFS}.
We then find the points on the intersection of the line $\lambda a+\rho b$ with $M$, in other words
the choices of $\lambda/\rho$ for which $f(\lambda a+\rho b)=0$.  This equation will have $\deg f$
solutions (with multiplicity) and for our purposes (which do not look at correlations between points) we
can use all of them as inputs $x_n$.
By a theorem of \citet{Shiffman1999},
they are distributed according to the pullback of \equationref{eq:defFS} to $M$.  
We then reweighted this distribution to get 
$d\mu_{ref}=d\mu_{\Omega}$, by multiplying by the ratio of this
form over the FS volume form.  In this way we avoided
introducing an additional geometric quantity not present in the actual CY
geometry.  This was important for the balanced metric computation of 
\citet{Douglas2008}, but for defining a loss function it is not
strictly necessary.

\subsection{ Implementation details }
\label{ss:impl}

Our code is written in Python 3 and uses TensorFlow, numpy and sympy.
Many of the routines are written twice, using sympy for generality and
then rewritten for our specific examples using numpy for efficiency.

To better organize the code, we defined 
a general class which implemented
operations such as sampling points, maintaining sets of points, integrating
over points, computing geometric tensors and the like, hiding details specific
to a particular construction of manifolds.  We put these details in the
{\tt Hypersurface} class.  One could write other classes for other manifold
definitions, for example hypersurfaces in toric varieties.

Although one might think that the existence of projective coordinates
eliminates the need to define coordinate patches, it is better
to use them.  This is because when we take derivatives
in \equationref{eq:defKahler}, 
if all of the coordinates are small (as is possible with
projective coordinates) we can lose numerical precision and potentially overflow.  For example,
after taking derivatives \equationref{eq:defKahler} of \equationref{eq:a1}, the norm of the
coordinates appears in the denominator.  This norm depends on the parameters
$h$ in \equationref{eq:a1}, and when we go to networks the denominator depends on the
weights in an even more complicated way.  For numerical
stability, it is better to assign the points to patches according to which
of their coordinates has the largest magnitude, and then normalize
this coordinate to $1$.  Then, constructing \equationref{eq:nform} on the hypersurface
leads to a second potential division by a small number. This is handled
by a further subdivision into subpatches, where a point is assigned to
a subpatch according to the largest magnitude of $|\partial f/\partial z^i|$.
The class structure {\tt Hypersurface} isolates these details from the
rest of the code, which need only know how to deal with several batches
of points (coming from the different subpatches). We also implemented a function to generate TensorFlow datasets which store the points on different patches and their patch information. They can be easily loaded and trained as in other neural network problems.

To integrate the code with TensorFlow, we constructed 
layers which implement \equationref{eq:defFK}.  
Of course we needed a special layer to compute the volume form \equationref{eq:MA} from 
the K\"ahler potential. This requires the computation of the complex hessian 
$\partial_i \partial_\bj K$. With some manipulations, it can be done directly using TensorFlow's backpropagation technique, which allows us to calculate the derivatives efficiently even with higher $k$s.

In a polynomial network with activation function $z\rightarrow z^2$, 
the weights in the front layers will have higher orders as the depth of the network increases, 
which will make the values of gradients unstable over each step. 
To solve this problem, we used the Adam algorithm \citep{Kingma2015} to train our network.
It computes an individual adaptive learning rate for each parameter in the network using the first and second moments, which smooths the training process. 

Adam and other gradient descent methods are first order, and converge slowly compared with
second order methods which use the Hessian.  
Newton's method is the simplest example.  If one starts close to a minimum, a second
order method will square the error $\epsilon$ with each iteration, and converge very quickly.  A rough estimate
is that $\epsilon\rightarrow \lambda_{1}|f'|\epsilon^2$ where $\lambda_{1}$ is the largest
eigenvalue of the inverse Hessian $\partial^2 f$.

The inverse Hessian has $D^2$ components in $D$ dimensions, and it is generally impractical to compute
it in a high dimensional parameter space.  To deal with this, one can estimate it from the gradient
computed at several points.  The L-BFGS algorithm \citep{DongC.Liu1989} does this using the gradients
computed in the previous iterations and is often the second order method of choice.  
It is not much used in practical 
machine learning, because the signal to noise ratio is generally not high enough to justify high precision
optimization, and it is not even available in standard Tensorflow.  However it is in an extension package
{\tt tensorflow\_probability} and it is available in our code.

Another method to solve the exploding/vanishing gradient problem in deep networks is batch normalization \citep{Ioffe2015}. This is done by standardizing the weights distribution in the training stage, which requires individual recenterings and rescalings of the components in the vector of sections. 
This is not allowed in our model, but one could still do a overall rescaling of the vector. We will implement this feature in the future.

\section{ Results }
\label{s:results}

In the present work we implement the algorithm just proposed,
study the dependence of accuracy and speed on the hyperparameters
(depth, widths, learning schedule), and propose reasonable values.
We intend to study interesting geometry and physics problems elsewhere.

Compared to previous work,
the main advantage of this approach is that it can describe arbitrarily
complicated metrics with structure on multiple scales.  To study this,
we considered several quintic CY manifolds with different parameterized
families of defining functions $f=0$.  We want to
vary the minimal length scale on which the metric will have structure,
and we want to scan through geometries with more or less symmetry.

The families we chose were
\begin{enumerate}
\item The Dwork quintics $f=0$ as in \equationref{eq:fermat}.  Equivalently, $f=f_1$ below with $\phi=0$.
\item A two parameter family with less symmetry,
\be \label{eq:def_f1}
f_1 = z_0^5 + z_1^5 + z_2^5 + z_3^5 + z_4^5 + \psi z_0z_1z_2z_3z_4 +\phi (z_3z_4^4 + z_3^2z_4^3 + z_3^3z_4^2 + z_3^4z_4)
\ee
\item Another two parameter family,
\bea \label{eq:def_f2}
f_2 = f_1|_{\phi=0} +
&\alpha & \bigg( 
 z_2  z_0^4 +  z_0  z_4  z_1^3  +  z_0  z_2  z_3  z_4^2 +  z_3^2  z_1^3 +   z_4  z_1^2  z_2^2 +  z_0  z_1  z_2  z_3^2 +  \\ 
 \nonumber
 & & z_2  z_4  z_3^3 +  z_0  z_1^4 +   z_0  z_4^2  z_2^2 +  z_4^3  z_1^2 +  z_0  z_2  z_3^3 +  z_3  z_4  z_0^3 + z_1^3  z_4^2 + \\
 \nonumber
 & &  z_0  z_2  z_4  z_1^2 +  z_1^2  z_3^3 +  z_1  z_4^4 +  z_1  z_2  z_0^3 +  z_2^2  z_4^3 +   z_4  z_2^4 +  z_1  z_3^4
\bigg).
\eea
\end{enumerate}

Whereas the Fermat quintic has $\IZ_5^4\times S_5$ discrete symmetry, taking $\psi \ne 0$
breaks this to $\IZ_5^3\times S_5$.
The generic CY in the $f_1$ family has $\IZ_5^2\times S_3$ discrete symmetry,
and the generic CY in the $f_2$ family has no discrete symmetry.\footnote{
We took the parameters $\psi,\phi,\alpha$ real so there is still a $\IZ_2$ 
complex conjugation symmetry.}
Thus specifying their metrics will require successively more data; we will
not try to quantify this dependence.  We should add that the functions
$f_1$ and $f_2$ were chosen before beginning our experiments.

A reasonable proxy for the smallest length scale is the distance in the space of defining
functions from $f$ to the nearest function $f_s$ which defines a
singular Calabi-Yau manifold, one for which $f_s=\nabla f_s=0$ has a simultaneous solution in $\IC\IP^4$.
As we briefly explain in the appendix, this is justified by looking at the generic form of a nearly singular
metric and how it depends on the coefficients of $f$.  A suitable definition of the distance is\footnote{
To clarify the notation in the denominator, a term $Z_i^5$ in $f$ gets the coefficient $5!/5!=1$, 
$Z_1^3 Z_2^2$ gets $3!\dot 2!/5!=1/10$, {\it etc.}.
}
\be \label{eq:dist_sing}
d^2_{sing}(f) = \frac{ \min_{Z\in M} \sum_{i=1}^5 \left|\frac{\partial f}{\partial Z^i}\right|^2 }
{ (1/5!) \sum_I (\prod_{i=1}^5 (I_i)!) |f_I|^2 }
\ee
These values are plotted as heat maps for $f_1$ in \figureref{fig:dist1} and $f_2$ in \figureref{fig:dist2}.

The accuracy of a solution will be measured as the norm of $\eta-1$ in the $L^1$ (MAPE), $L^2$ (RMSE)
and $L^\infty$ (MAX) error.  In machine learning, one always checks the error twice, for the sample
used in optimization (training error) and on another set
of points chosen independently (testing error).
We followed this practice, in part because it is the default in Tensorflow/Keras, and also to get
a conservative error estimate.  As we discuss below the difference will be meaningful.

A particular run of the algorithm will depend on various ``hyperparameters'': 
\begin{itemize}
    \item The depth of the network and width of the layers.
    \item The number of points $N_p$ for Monte Carlo integrations.
    \item The batch size $N_b$ for stochastic gradient descent.
    \item The optimization algorithm and training schedule (number of gradient descent steps, learning rate, {\it etc.}).  
\end{itemize}

Practically, the most stringent constraints on these parameters are 
memory limitations.  The largest memory items in a network are usually 
the intermediate results,
which for width $W$ are matrices of size $N_b\times W$,
with each item taking $4-16$ bytes depending on precision and 
whether complex numbers are needed.
Our 32 GB GPUs allowed batches of 100000 points but this was cut to
20000 points for
the largest networks (see \tableref{tab:summary}).

For a single layer network directly implementing  
the Fubini-Study metrics \equationref{eq:defFS}, $W$ is the total number of parameters.
This grows rapidly with $k$ and our memory limited us to $k \le 4$.\footnote{
With additional programming effort one could probably do $k\le 6$.
By using minibatches one could go higher, but this would require 
replacing our L-BFGS optimization, perhaps as in \citet{Berahas2016}.}
By contrast one can run 5 layer and even deeper FNN's.

Out of the networks which can be run, one wants a choice which
works robustly (in nearly all cases) with minimal error.
Since all metrics of a given $k$ are Fubini-Study metrics
, we should compare with the minimal error in 
this class.
Assuming $M$ is nonsingular, this best possible 
error will decrease exponentially in $k=2^{\mbox{depth}}$.
As explained in \citet{Donaldson2009} this follows because we are making a
polynomial approximation to a $C^\infty$ function.  The corresponding statement
in Fourier space may be more familiar (the Paley-Wiener theorem).

It is not {\it a priori} clear to us that this will be the case for our networks.
One might hypothesize that the error is controlled instead by the total number of parameters in
the network (denote this number as $P$), the width
(as found in \citet{Golubeva2020}),
or some other property. 
So far as we know, work on neural methods for PDE 
\citep{Weinan2018,Grohs2019,Muller2019}
has not led to a clear conjecture about this, but
it is an interesting question to study.

We next discuss the numbers of total points $N_p$ and batch size $N_b$.
In machine learning, one usually takes $N_b<N_p$ both for computational practicality and
to get a noise term in the gradient, parametrically of magnitude $1/\sqrt{N_b}$.
Empirically this leads to models which generalize better,
for reasons which are not completely sorted out but which include the following.
First, ML loss functions are usually non-convex, and
noise helps the optimization to escape local minima.
This turns out to be the case for our loss functions as well.
Second, noise favors finding wide minima (with small
second derivatives or even flat directions), and
there are statistical arguments that these will generalize better.  
While our problem is not statistical, since we are using sampling
in our computations, this point might be relevant.  However we
did not find any evidence that decreasing $N_b$ ever improves our results.

A much discussed point in the theory of machine learning is the role of overparameterization.
In general, once the number of parameters exceeds the total dimension of the data being fit,
one expects to be able to completely fit (interpolate) the data.
The optimization problem is easier in this regime,
which is a significant advantage.
But according to the usual dogma of statistics and of numerical analysis, as one uses more
parameters, the extra fitting ability will fit noise, and the accuracy on testing data will decrease.  One might have thought that an
overparameterized model would not be able to generalize. 

Surprisingly, overparameterized models can generalize well.
While the reasons for this are still being debated, it seems to be accepted
that the dogma we cited is not correct in this regime, with 
works such as \citet{Zhang2017,Belkin2019} making many observations
which contradict it.  It still might be that this paradox can be resolved
within current theory.  For example,
a traditional way to deal with overfitting
is to introduce a regularization term in the loss function, such as the sum of the squares of the weights,
which favors simpler models \citep{bishop2013pattern}.
A popular hypothesis is that the choice of initial conditions and optimization algorithm in deep learning produces an implicit regularization term.

Previous works on CY metrics studied highly symmetric metrics with 
few parameters, so this issue did not arise.  But
we will encounter this issue,
now in a mathematically controlled setting.
Thus 
in \S \ref{ss:interp} we also looked at the case $N_p<P$ to see if the situation
is similar to that for ML.

\subsection{Experiments and Observations}
\label{ss:experiments}

As we explained above, a holomorphic network with width less than the number of sections at a given layer
is not able to represent the Fubini-Study metrics at that $k$, which might lead to problems.  In practice
we found that these results were very sensitive to the initial conditions for the gradient descent.
The bihomogeneous networks were not very sensitive and produced better results, so we restricted the
rest of our study to these.

The convergence of some example networks is shown in \figureref{fig:train_curve}. 
Adam seems to converge slowly near the minima, so
we also introduced a second stage of training using L-BFGS and $N_b=N_p$.
This solved the speed problem, converging in a matter of minutes. 

To understand the magnitude of the errors, it is useful to estimate the optimal error as
a function of $k$.  We did this following an observation of \citet{Headrick2013}.
They found for the Dwork quintics that the error \equationref{eq:Lquad} went as
$E \propto C^{-k}$, with $C\sim 8$ for the Fermat quintic and varying with $\psi$.
The fit was already good at low $k$, so by fitting this formula with low values of $k$,
we can estimate the best possible error as a function of $k$ and the parameters of the CY.
Thus we optimize over the entire space of metrics \equationref{eq:defFS} for $k=2,3,4$ and do the linear
fit
\be
\log \mbox{error} \sim C_0 + C_1 k .
\label{eq:est_error}
\ee
This leads to estimated optimal errors, denoted in the plots as {\tt est8} for $k=8$. We then did sweeps through the parameters of the three families of quintics and compared the results with these optimal errors.

Overall, the network configurations have the largest effect on accuracy. 
We experimented with networks 50\_50\_1, 70\_70\_70\_1, 300\_300\_300\_1 and 500\_500\_500\_500\_1, etc., corresponding to $k = 4,8,16$ with various total number of parameters (see \figureref{fig:mse1}, \figureref{fig:mse2} and \tableref{tab:summary}). 
In general, $k$ plays a more important role here, which agrees with the previous results of
\citet{Headrick2013}: The accuracy of the network models are able to reach approximately the same magnitude
as those of the FS models with the corresponding $k$ in all cases, estimated by \equationref{eq:est_error}.
The dependence on the number of parameters is more complicated.
It does not have a significant independent effect for the more symmetric quintics, 
but it does for those defined by \equationref{eq:def_f2}, as we discuss in \S \ref{ss:interp}.
For example, the $k=8$ FS models with 245025 real parameters are predicted to be
around a factor of 100 better MSE than $k=4$. However, the 70\_70\_70\_1 network was able to 
achieve the same accuracy with only 11620 real parameters for the more symmetric manifolds $f_1$, 
and increasing the number of parameters did not seem to improve it.  
But for $f_2$, a wider 300\_300\_300\_1 network with 187800 real parameters 
significantly improved the training accuracy in most cases,
although much of this was overfitting (especially for $f_1$)
as one can see in 
\figureref{fig:overfit1}.\footnote{
This is another motivation to implement minibatched L-BFGS as mentioned in
\S \ref{s:results}.
}  
Still, this model was able to
attain the FS $k=8$ accuracy in all cases.
While the $k=8$ FS metrics are simpler than such a network, its
memory requirements (at least for a
Tensorflow implementation) are challenging, so reproducing its
accuracy is of real practical value.

There is also a strong though not exact relation between the distance to the singularity and the error, supporting the idea that the ratio
of length scales controlled by this is the dominant parameter. 
This relation is shown in \figureref{fig:mse1} and \figureref{fig:mse2} and summarized in \tableref{tab:summary}, where we quote an average error for "less singular" and 
"more singular" CYs with distance to the discriminant locus less than (greater than) 0.1.
The distance also affects the total number of parameters needed to reach the best accuracy.
The 70\_70\_70\_1 network shows similar performance as the 300\_300\_300\_1 one for less singular $f_2$s
(distance greater than 0.15), in comparison to other cases mentioned above.

In scanning through a family of manifolds,
training could be made significantly faster by initializing the network with
the weights optimized for a nearby parameter value.  Details of how to do this and
the appropriate learning rates and times can be found in the accompanying software.

For completeness, we also 
checked the idea that the metrics on the CY manifolds defined by  
\equationref{eq:def_f1} and \equationref{eq:def_f2} must be described by non-symmetric models
as follows.  
We computed K\"ahler metrics using the code developed by 
\citet{Headrick2013}, which assumed the discrete symmetry $S_5\times Z_5^4$,
and computed the deviation from Ricci flatness for all
three classes of CY defined in \S \ref{ss:geom}.
For $k=4,8$ and the symmetric CYs \equationref{eq:fermat},
this was roughly the same as for our 
models.  But as one would expect, the errors for \equationref{eq:def_f1} and \equationref{eq:def_f2}
with $\phi,\alpha\ne 0$ were large, of order $0.1-0.01$, even for high $k$.

Some other methods we tested include $\ell_2$ regularization, using 64 bit networks and changing loss functions.
Most of them did not show a significant impact on our results, 
but in some cases, it could be helpful to add 0.1*MAX error to the loss function in the early stage of training 
to prevent getting stuck in a bad local minimum, especially for a deep network and a complicated manifold.

Lastly, we made several attempts to find a model which significantly improves on 
the performance, including the five layer 300\_100\_100\_100\_1 and 500\_500\_500\_500\_1
models, and also using the values of the $k=2$ or $k=4$ sections as inputs
to less deep networks. 
We found that the optimization was much less robust, often leading to
bad local minima.  In the second L-BFGS stage,
memory limitations restricted us to $N_p\sim 2000$.
While some of our runs had up to 10 times better accuracy, 
these successes were not reliable.  
Thus for math physics applications we suggest at present
using the four layer $N\_N\_N\_1$ networks with $N\sim 100$.

We believe that further development could improve on this,
and our GitHub site will have a benchmark script with a set
of test examples and a leaderboard page where we will post new
state of the art results.

\subsection{ More theoretical issues }
\label{ss:interp}

We turn to discuss theoretical issues for which these results might be relevant.  One question we raised earlier is whether the accuracy is
controlled more by $k$ or by the number of parameters.  While $k$ controls the 
smallest length scale on which one can vary the metric, it
might also be that a complicated metric contains many features at
a variety of scales,
which require many parameters to represent.  Comparing
our results for the 70\_70\_70\_1 network and the 300\_300\_300\_1
network suggests that both factors are in play.  Whereas both networks
attained the maximal accuracy for the simpler and
more symmetric hypersurfaces $f_1=0$, for the more complicated 
hypersurfaces $f_2=0$ and the more singular cases of these,
only the larger network attained this accuracy.  This suggests
that there is a ``complexity'' factor which deserves to be
quantified and understood.

In theory of ML, the overparameterized regime is usually defined not in terms of an inequality between $N_p$ and $P$, but rather as the regime in which the model has enough parameters to fit an arbitrary set of $N_p$ observations
\citep{Zhang2017}.  To locate this threshold in our case, we did runs computing
the best $k=4$ FS metric (with $P=4900$ parameters) and with a range of
$N_p$ values.  We found that this model could
achieve roughly zero training loss for $N_p\le 2500$, while
the (MSE) testing error had a significant dependence on $N_p$,
consistent with $1/\sqrt{N_p}$.
Already for
$N_p=5000$ the training error was comparable to that for larger $N_p$.
Following \citet{Zhang2017},
we also trained models with ``random labels,'' here produced by
shuffling the $\eta$ values between the different points, and found
the same behavior of training error, along with large testing error.
This case can be compared with a random feature model which is solvable
in the large $N_p,P$ limit, as we will discuss elsewhere.

\section{ Conclusions and further directions }

We developed and tested software to compute Ricci flat metrics
on Calabi-Yau hypersurfaces in projective space.  Using it, one can get
results with $\sim 0.1\%$ absolute error on CY$_3$ manifolds
with no symmetry.
It is based on the standard Tensorflow/Keras
platform and can be easily adapted to handle more
general K\"ahler manifolds.   

It would be interesting to look at the Laplacian
and hermitian Yang-Mills equations and the 
other applications
considered in previous work.
Similar techniques could be used to represent other geometries, 
as we intend to discuss elsewhere. 

The Ricci flat K\"ahler metric problem is one of the
better understood problems in nonlinear PDE, with no 
boundary conditions to complicate the discussion,
so it might serve as a good testbed for numerical methods.
It would be interesting to try other deep learning PDE methods
such as \cite{sirignano2018dgm}.
And as we discussed in \S \ref{ss:experiments},
one can get a well motivated estimate for the
maximal accuracy which can be obtained for a given depth
network.  This helps in studying how
the accuracy depends on hyperparameters, as we discussed in
\S \ref{ss:interp}.
It would be interesting to
have more specific theoretical predictions for this dependence.

\acks{This project was begun in the summer of 2019 at Stony Brook University 
in collaboration with Tudor Ciobanu.  Tudor was a very promising first
year graduate student, and his untimely passing fills us with deep sorrow.
We hope this work can add to his legacy.

We thank the authors of \citet{anderson_moduli-dependent_2020} for early discussions and informing us about their work.

We thank Steve Skiena
and the Stony Brook AI Institute for the use of their GPU servers.

MRD would like to thank Shing-Tung Yau for discussions,
for reading the manuscript and for general support.  He thanks Steve
Zelditch for discussions on K\"ahler geometry and comments on \citet{Douglas2020}.
He also thanks many people for conversations about ML, and especially
Sanjeev Arora, David McAllester, Andrea Montanari and Christian Szegedy.}

\bibliography{msml2021.bib}

\newpage
\appendix
\section{Plots and tables}
\begin{figure}[htbp]
\floatconts
    {fig:dist}
    {\caption{Distance to singular CY as function of $\psi,\phi$ in \equationref{eq:def_f1} (Left) and $\psi,\alpha$ in \equationref{eq:def_f2} (Right)}}
    {%
      \subfigure{
          \label{fig:dist1} 
          \includegraphics[width=0.45\linewidth]{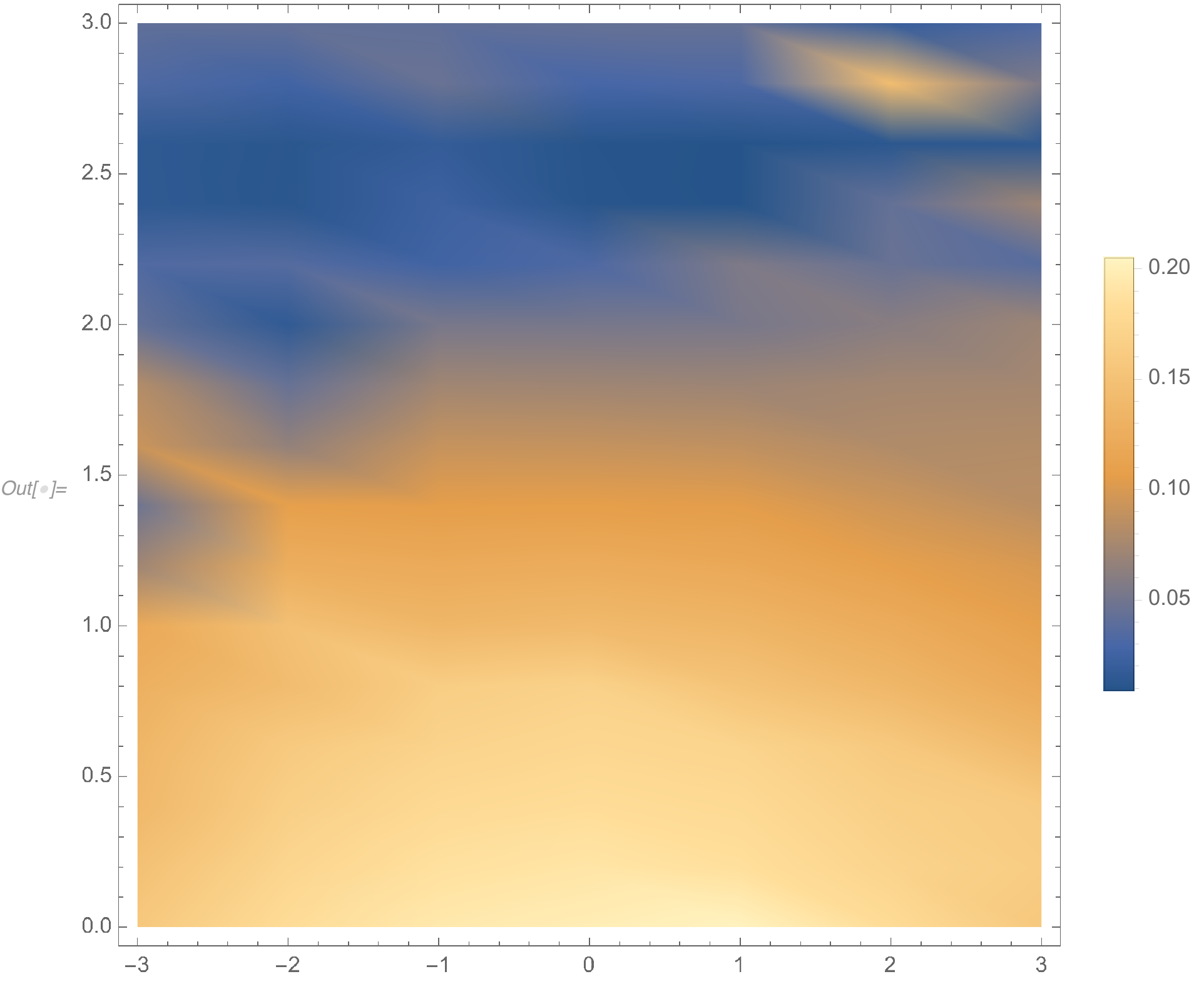}
      } \qquad
      \subfigure{%
          \label{fig:dist2}
          \includegraphics[width=0.45\linewidth]{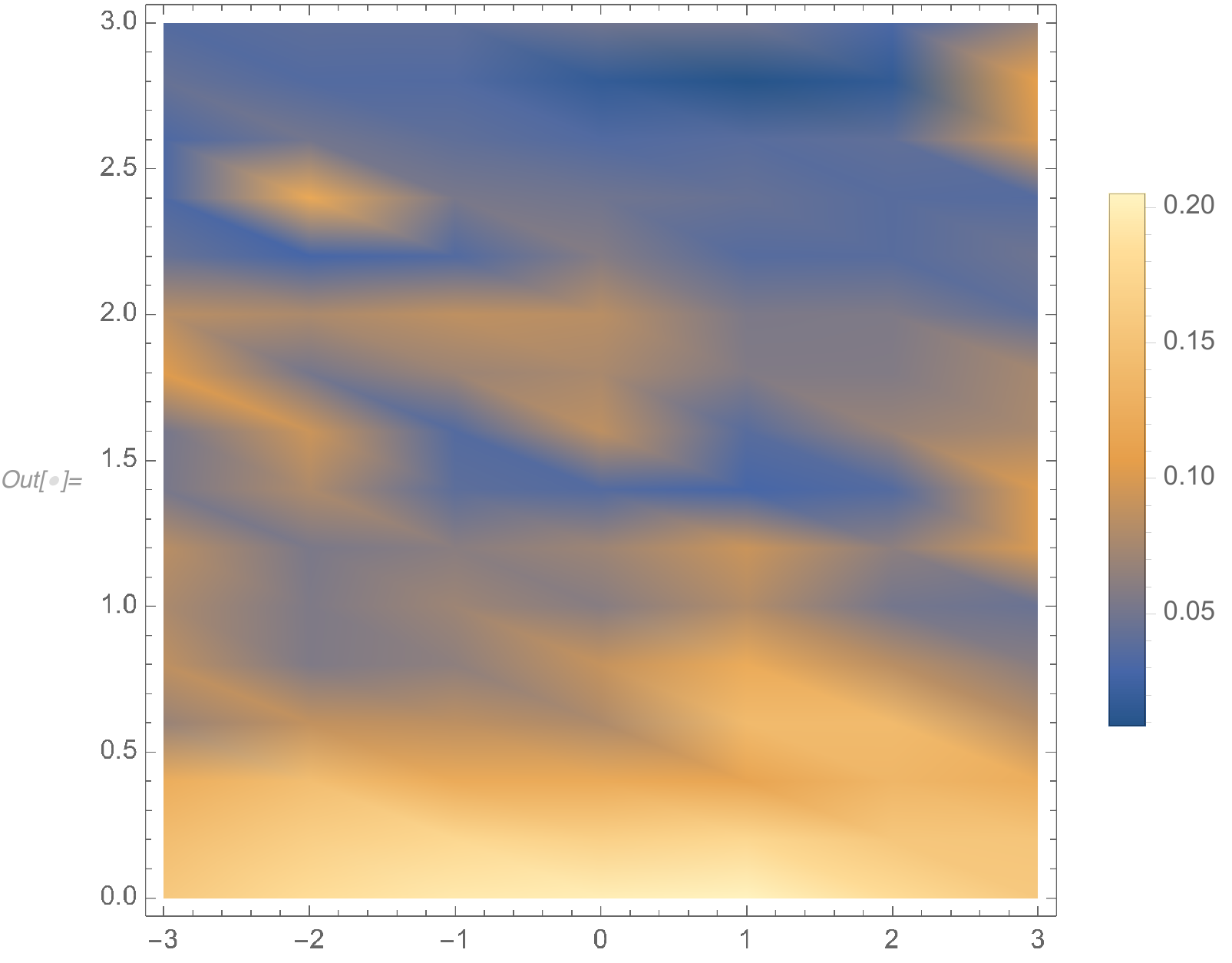}
      } 
    }
\end{figure}

\begin{figure}[htbp]
\floatconts
    {fig:train_curve}
    {\caption{The training curves for \equationref{eq:fermat} with $\psi$ = 0.5, trained with Adam optimizer and MAPE loss. The data for k2\_500\_500\_500\_1 was recorded every 10 epochs.}} 
    {\includegraphics[width=0.8\textwidth]{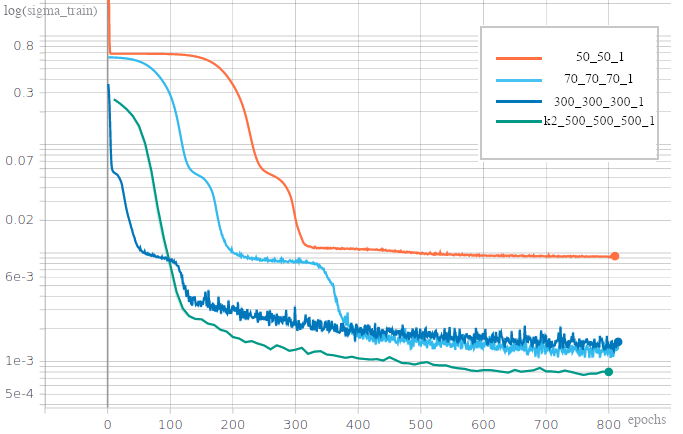}}
\end{figure}

\begin{figure}[htbp]
\floatconts
    {fig:mse1}
    {\caption{Mean testing $(\eta-1)^2$ function of $d_{sing}$ in f1 \equationref{eq:def_f1} with k = 2, 3, 4, the extrapolated k = 8, and three neural network models}}
    {\includegraphics[scale=0.4]{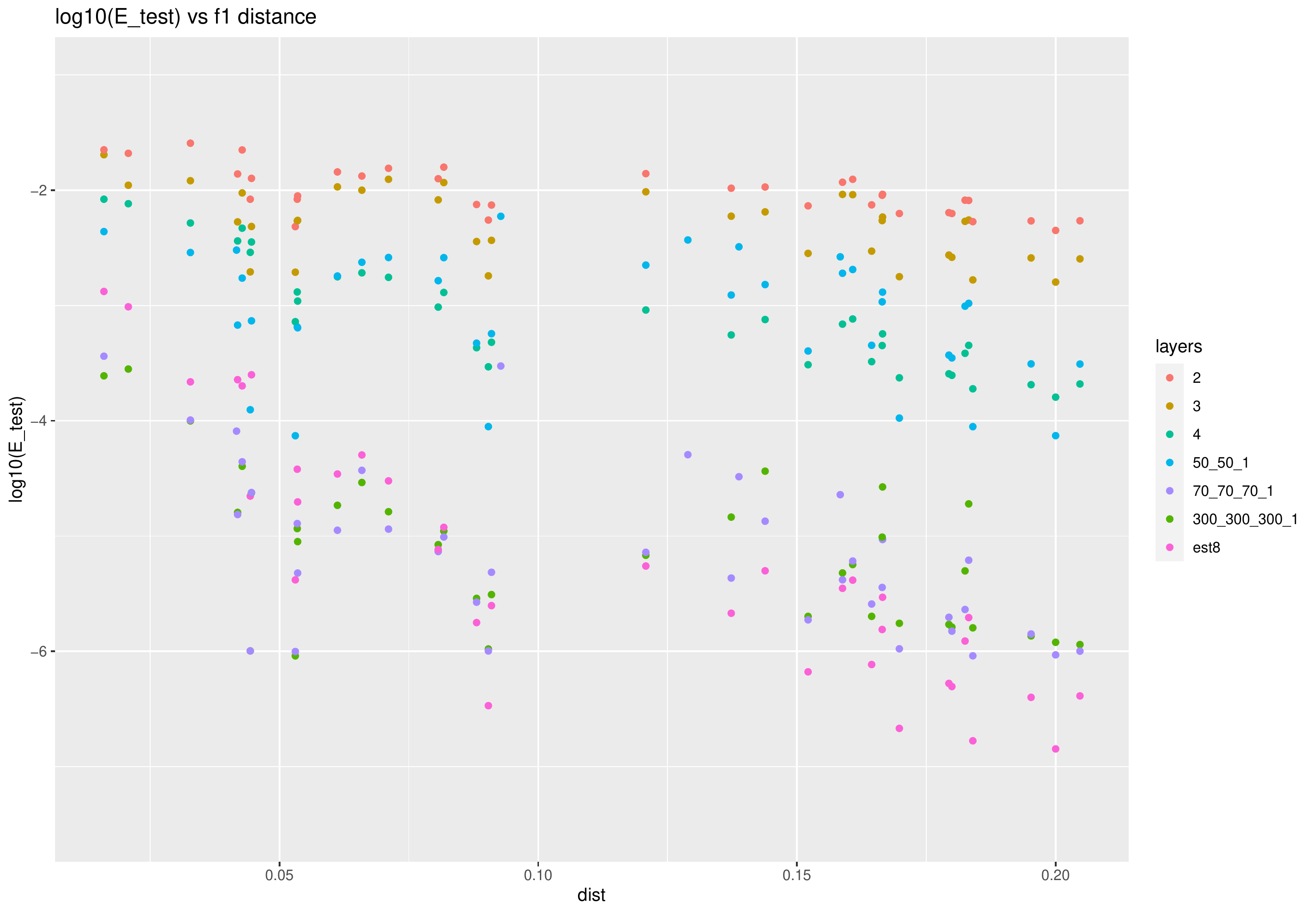}}
\end{figure}

\begin{figure}[htbp]
\floatconts
    {fig:mse2}
    {\caption{Mean testing $(\eta-1)^2$ function of $d_{sing}$ in f2 \equationref{eq:def_f2} with k = 2, 3, 4, the extrapolated k = 8, and three neural network model}}
    {\includegraphics[scale=0.4]{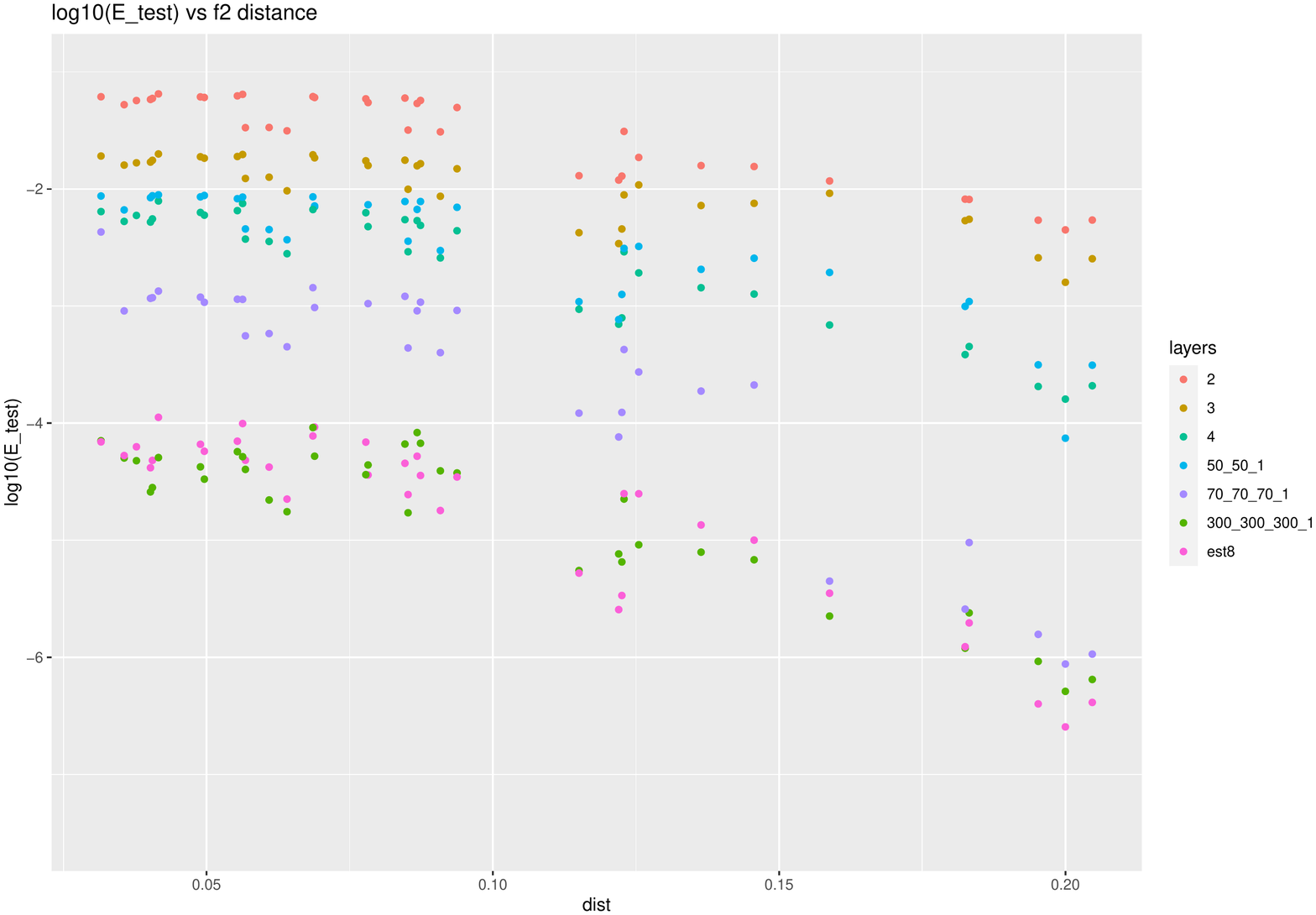}}
\end{figure}

\begin{figure}[htbp]
\floatconts
    {fig:overfit}
    {\caption{The overparameterization of L-BFGS}}
    {%
        \subfigure[log(E\_test) vs log(E\_train) with L-BFGS]{%
            \label{fig:overfit1}
            \includegraphics[width=.45\linewidth]{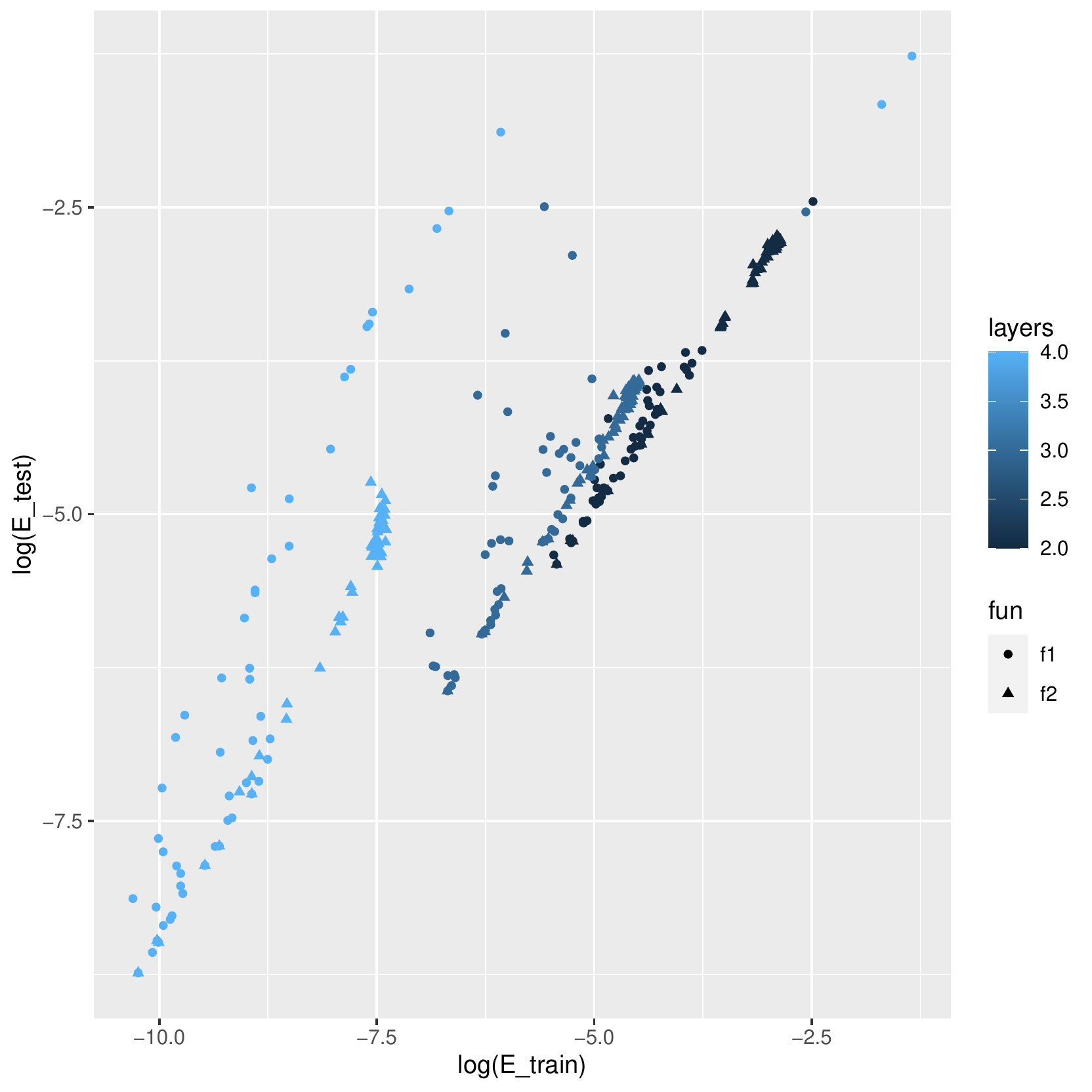} 
        }\qquad
        \subfigure[f1 log(E\_test) vs dist for $300\_300\_300\_1$, L-BFGS and different numbers of point groups (each with five points).  The "NA" entry is the extrapolated FS accuracy at $k=8$.]{%
            \label{fig:npoints}
            \includegraphics[width=.45\linewidth]{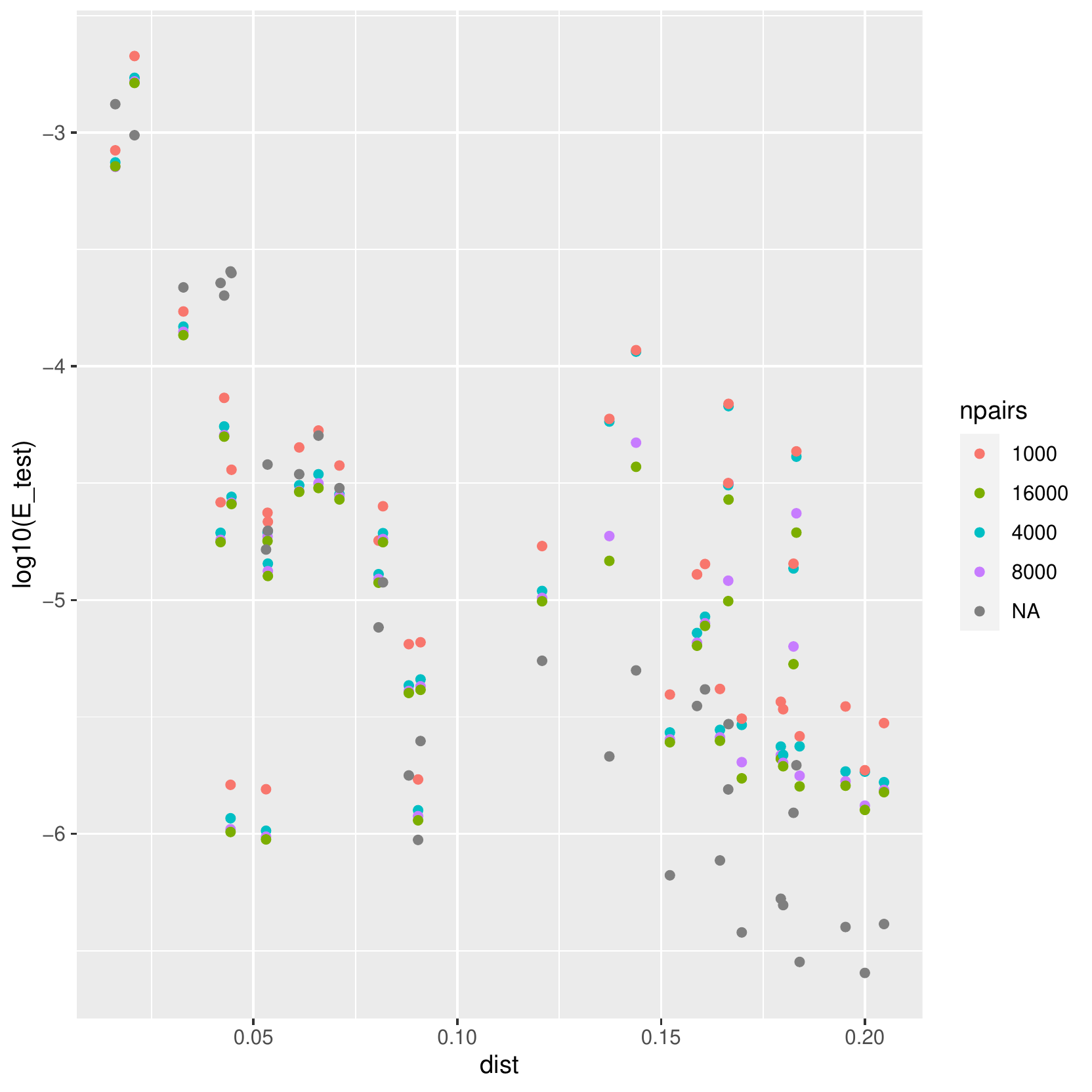} 
        }
    }
\end{figure}

\begin{table}[htbp]
\floatconts
    {tab:summary}
    {\caption{A summary of the average $\log_{10}(\textrm{E\_train})$ and  $\log_{10}(\textrm{E\_train})$ over runs of different distances}}
    {%
    \begin{tabular}{lrrrrcc}
      \hline
      func & distance & layers & n\_params & n\_pts\_lbfgs & $\langle \log_{10}(\textrm{E\_train})\rangle$ & $\langle \log_{10}(\textrm{E\_test})\rangle$ \\ 
      \hline
      f1 & $<$0.1 & 50\_50\_1 & 3800 & 100000 & -3.07 & -3.00 \\ 
      f1 & $>$=0.1 & 50\_50\_1 & 3800 & 100000 & -3.17 & -3.14 \\ 
      f1 & $<$0.1 & 70\_70\_70\_1 & 11620 & 100000 & -5.02 & -4.86 \\ 
      f1 & $>$=0.1 & 70\_70\_70\_1 & 11620 & 100000 & -5.45 & -5.40 \\ 
      f1 & $<$0.1 & 300\_300\_300\_1 & 187800 & 80000 & -5.94 & -4.89 \\ 
      f1 & $>$=0.1 & 300\_300\_300\_1 & 187800 & 80000 & -5.78 & -5.38 \\ 
      f2 & $<$0.1 & 50\_50\_1 & 3800 & 100000 & -2.22 & -2.18 \\ 
      f2 & $>$=0.1 & 50\_50\_1 & 3800 & 100000 & -3.03 & -3.01 \\ 
      f2 & $<$0.1 & 70\_70\_70\_1 & 11620 & 100000 & -3.15 & -3.01 \\ 
      f2 & $>$=0.1 & 70\_70\_70\_1 & 11620 & 100000 & -4.68 & -4.62 \\ 
      f2 & $<$0.1 & 300\_300\_300\_1 & 187800 & 20000 & -5.86 & -4.37 \\ 
      f2 & $>$=0.1 & 300\_300\_300\_1 & 187800 & 20000 & -6.11 & -5.48 \\ 
       \hline
    \end{tabular}
    }
\end{table}

\section{ Reviews of the concepts}
\label{appendix:reviews}

\subsection{ Complex and K\"ahler geometry }
\label{appendix:kahler}

This is a very brief introduction meant for readers familiar with real differential geometry
and the concepts of
manifolds, tensor fields, differential forms, Riemannian metrics, and curvature.

A complex manifold $M$ is a topological space which is locally similar to the $d$-dimensional
space $\IC^d$.  One can define it in terms of coordinate patches $U_\alpha$ whose union is $M$,
and coordinate maps $Z_\alpha$ which are complex diffeomorphisms between $U_\alpha$ and a contractible
subset of $\IC^d$.  As a linear space, $\IC^d\cong\IR^{2d}$, with coordinates $Z^i=X^i+iY^i$,
and thus every complex manifold is
also a real manifold.  But the complex coordinate systems provide additional structure, summarized by
a linear operator $J_x$ at each point $x\in M$, which turns
a tangent vector corresponding to a small motion in the $\Re Z^i$ direction (and corresponding to
$\p/\p \Re Z^i$) into the vector corresponding to $\Im Z^i$.  These operators combine to a linear
tensor operator which maps the tangent space $TM\rightarrow TM$, and can be denoted in tensor notation
as $J^i_j$.  Complex geometry can also be phrased as a ``special geometry'' in which operations such
as the covariant derivative preserve the complex structure, so $\nabla_i J^k_l=0$.  From this point of
view, all of the concepts of real geometry we listed above
have the same primary definitions in complex geometry, but they
satisfy additional constraints.

In complex geometry one often uses the ``bar'' notation for complex conjugation and tensors.
Complex conjugate coordinates are denoted $(Z^i)^{*}$ or interchangeably $\bar Z^\bj$.
Thus, a tangent space $T_xM$ is a direct sum of a complex tangent space $T_{\IC,x}M$ with
coordinate basis $(\p/\p Z^1,\ldots,\p/\p Z^d)$
and a complex conjugate tangent space with coordinate basis $(\p/\p \bZ^1,\ldots,\p/\p \bZ^d)$.
Here $\p/\p Z^i=\frac{1}{2}(\p/\p X^i-i\p/\p Y^i)$, with the constant chosen so that 
$\p Z^j/\p Z^i =\delta_i^j$.  We have $\p Z^j/\p \bZ^i =0$ and 
a holomorphic function $f$ is defined as one for which $\p f/\p \bZ^i =0 \,\forall i$,
generalizing the definition in one complex dimension.

The Euclidean metric $ds^2 = (dX^1)^2 + (dY^1)^2 + \ldots + (dX^d)^2 + (dY^d)^2$
becomes $ds^2 = dZ^1 d\bZ^1 + \ldots + dZ^d d\bZ^d$,  This is written in tensor notation as
$ds^2 = g_{i\bj} dZ^i d\bZ^\bj$ with $g_{i\bj}\equiv \delta_{i,\bj}$ with components $1$
if the indices agree ($i=\bj$) and zero otherwise.  A general Riemannian metric can be written
in this notation and could 
also have terms $g_{ij} dZ^i dZ^j$, $g_{\bi\bj} d\bZ^\bi d\bZ^\bj$ and 
$g_{\bi j} d\bZ^\bi dZ^j$.  But we will only consider hermitian metrics, for which the metric
tensor is a positive definite hermitian matrix.  In complex notation this requires 
$g_{ij}=g_{\bi\bj}=0$ and $g_{\bi j}=g_{i\bj}^*$.

After $\IC^d$, the next simplest example of a complex manifold is complex projective space $\IC\IP^d$.
This can be defined in terms of patches, but the shortest and clearest definition is as
a quotient $\IC\IP^d\equiv \{\IC^{d+1} - \vec 0\}/\sim $.  
Let the complex coordinates of $\IC^{d+1}$ be $(Z^1,Z^2,\ldots,Z^{d+1})$,
then $(Z^1,Z^2,\ldots,Z^{d+1}) \sim (\lambda Z^1,\lambda Z^2,\ldots,\lambda Z^{d+1})$
for every $\lambda\in\{\IC-0\}$.  One can cover $\IC\IP^d$ by coordinate charts $U_\alpha$,
where $U_\alpha$ is defined by choosing the representative of $\sim$ in which $Z^\alpha=1$.
The space $\IC\IP^1$ is topologically identical to $S^2$.
It is $\IC$ with an additional point at ``infinity,'' or the Riemann sphere.  The cases $d>1$
are not homeomorphic to spheres.
Like the spheres, each complex projective space has a maximally symmetric metric, called the
Fubini-Study metric.  For $\IC\IP^1$ it is isomorphic to the usual round metric on $S^2$.

This metric is best defined by first defining the more general concept of K\"ahler metric.  This is a metric
which locally (so, in each patch $U_\alpha$) is the second derivative of a real function $K_\alpha$ on $U_\alpha$,
\be \label{eq:defKahler2}
g_{i\bj} = \frac{\p^2}{\p Z^i\p \bZ^\bj} K_\alpha .
\ee
The K\"ahler potential for the Euclidean metric on $\IC^d$ is $K=\sum_i |Z^i|^2$.
These metrics are hermitian, but have many further special properties and simplifications.

Now, the Fubini-Study metric on $\IC\IP^d$
can be defined by the K\"ahler potential
\be \label{eq:defFS}
K = \log \sum_{i,\bar j=1}^{d+1} h_{i\bar j} Z^i \bar Z^{\bar j} ,
\ee
where $h$ is a $(d+1)\times (d+1)$ positive definite hermitian matrix $h$.
Note that this is not a function on $\IC\IP^d$, because it is not invariant under
the equivalence relation $\vec Z\sim\lambda\vec Z$.  However since it changes 
by addition of a term whose second derivative \equationref{eq:defKahler2} vanishes, it defines a metric
on $\IC\IP^d$.  One can also think of this as specifying a different (yet compatible) function in each
patch $U_\alpha$.
One gets the same metric on $\IC\IP^d$ for any choice of $h$.
This is because one can always find a linear change of coordinates $Z^i\rightarrow L^i_j Z^j$
which turns $h$ into the identity matrix.  Thus
one usually speaks of ``the'' Fubini-Study metric on $\IC\IP^d$.

The main object of our studies here will be hypersurfaces in $\IC\IP^d$.
A hypersurface in $\IC^{d+1}$ is the set of solutions to an equation $f=0$, 
as in \equationref{eq:fermat}.  If $f$ is a homogeneous polynomial,
the quotient by the relation $\sim$ above also makes sense for the hypersurface,
thus defining a hypersurface in  $\IC\IP^d$, which is a $d-1$-dimensional manifold
if the condition \equationref{eq:disc} is satisfied at each point.  While this construction only
produces a small subset of the possible complex manifolds, these already exhibit a 
great diversity of behavior.  The specific choice in \equationref{eq:fermat} was made to get
manifolds with Ricci flat metrics, as explained in the literature.

As in real geometry, given a map $M\rightarrow N$, 
one can pull back a metric on $N$ to get a metric on $M$.  One can check that for K\"ahler
geometry, this can be done by first restricting the K\"ahler potential $K$ from $N$ to $M$ and then
applying \equationref{eq:defKahler2} on $M$.  To relate derivatives on $M$
to derivatives on $N$, one uses $\partial_i f|_M=0$ to solve for one component
of $\partial_i|_N$ in terms of the others.  This defines a projection matrix
such that $\partial_i|_M = L_i^{i'}\partial_{i'}|_N$,
\be\label{eq:defLmatrix}
L^{i'}_i = \begin{cases} \delta^{i'}_i, \qquad & 1\le i,i'\le n, \\
-(\partial f/\partial Z^i)/(\partial f/\partial Z^{n+1}), \qquad & i'=n+1
\end{cases}
\ee

Thus we can regard \equationref{eq:defFS} as defining a family of metrics
on a hypersurface, depending on the $(d+1)^2$ real parameters of $h$.
Whereas these were equivalent on $\IC\IP^d$, since the linear transformation 
$Z^i\rightarrow L^i_j Z^j$ will generally not preserve the function $f$, these metrics are
generally distinct on $M$.  This gives us a parameterized family of metrics on $M$,
but of too low dimension for our purposes.  The next subsection will explain how we can get
larger families.

While we will not review curvature in detail, the formulas which
determine the connection and curvature in terms of the metric are much simpler in K\"ahler
geometry.   We quote the Ricci curvature,
\be \label{eq:KRicci}
R_{i\bj} = \frac{\p^2}{\p Z^i\bar\p \bZ^\bj} \log \det_{i,\bj} g_{i\bj} ,
\ee
which can be used to justify \equationref{eq:MA}.

\subsection{ Line bundles and the embedding method }
\label{appendix:embedding}

Following \citet{Donaldson2009}, most work on numerical Calabi-Yau metrics
represents the metric using an embedding by holomorphic sections of a very ample line bundle $\CL$.  This embedding is a map into a linear space, analogous to
spectral embeddings such as the ``Laplacian eigenmap'' construction, but with the great advantage that the map has a simple exact form.
Let us briefly review it.

A section of a holomorphic line bundle is locally a holomorphic function.
To define it globally, we define the line bundle by choosing patches $U_\alpha$ on the manifold $M$ and holomorphic transition
functions $f_{\alpha\beta}$ on the overlaps of patches $U_\alpha\cup U_\beta$ satisfying the
consistency conditions $f_{\alpha\beta}f_{\beta\gamma}f_{\gamma\alpha}=1$.  A section $s$ is then a holomorphic
function $s_\alpha$ on each patch satisfying $s_\alpha=f_{\alpha\beta}s_\beta$.  Now this is ambiguous
considered as a function, because we can always multiply by a set of holomorphic functions $\lambda_\alpha$
defined on each patch, taking $s_\alpha\rightarrow \lambda_\alpha s_\alpha$.  But the ratio of a pair of sections is
unambiguous.  This can be rephrased as the statement that a vector of $N$ sections is an unambiguous map $\iota$
from $M$ to $\IC\IP^{N-1}$, since $\lambda_\alpha$ acts to rescale the entire vector.
The very ample condition then states that the map $\iota$ is an embedding.

Consider the quintic hypersurface defined by \equationref{eq:fermat}.  In this case,
one can show that all holomorphic line bundles extend to the ambient space
$\IC\IP^4$.  These are parameterized by an integer $k$ and denoted $\CO_{\IC\IP^4}(k)$.
For $k\ge 0$ they have holomorphic sections, which are precisely the homogeneous
polynomials of degree $k$ in the coordinates $Z^i$.  These form
a linear space which is denoted $H^0(\CO_{\IC\IP^4}(k))$.
Take $k=1$ for example.  $H^0(\CO_{\IC\IP^4}(1))$ 
is the space of linear polynomials
in the homogeneous coordinates $Z^i$.
Its dimension, denoted $h^0(\CO_{\IC\IP^4}(1))$, is $5$.
Similarly $h^0(\CO_{\IC\IP^4}(k))=\binom{k+4}{k}$.

Because a section is locally just a function, it can be restricted to
a submanifold, thus defining the bundles $\CO_{M}(k)$.
This map is not injective -- a section proportional to the defining 
polynomial $f$ will restrict to zero.
This is nontrivial for
$k\ge 5$ -- for example for $\CO_M(5)$, the sections are fifth
degree polynomials with a single redundancy: if we add $f$ to
the section with an arbitrary coefficient, since $f=0$ on $M$,
we do not change the section.
To get a complete and nonredundant basis,
one needs to take this into account.  But since our
neural networks will generate spaces of sections which are not complete 
and can be redundant, we will not bother to take this quotient.

In general,
the representation of a manifold by an embedding has advantages and disadvantages.
Two disadvantages are that it can be hard to construct explicit coordinate charts, and
the embedding is an additional structure which may or may not be well suited to
the problem at hand.  In our case, we will not need explicit coordinate charts;
the only global operation we will need is to do integrals on $M$, and this can
be done by Monte Carlo (sampling points), as in \citet{Douglas2008}.

The second problem is mitigated if one can find a canonical embedding, 
determined by the intrinsic geometry of $M$ and not involving other choices.
This is indeed the case here: if we use a complete basis of
sections, which we can do because the basis is finite dimensional, the embedding
depends only on the choice of line bundle.  

The embedding representation gives us
a natural family of metrics: we choose a family of metrics on the
embedding space, and the pullback to $M$ gives us a family on $M$.
For an embedding in $\IR^N$, we could take the Euclidean metrics
$g_{ij} dx^i dx^j$ parameterized by a symmetric matrix $g_{ij}$.
While on $\IR^N$ these are all related by linear change of coordinates,
once we pull back to $M$ this generally provides a family of distinct metrics.

In the case at hand, the natural family of K\"ahler metrics is the family of
Fubini-Study metrics on complex projective space.  
As we discussed in \appendixref{appendix:kahler}, using the original
ambient space $\IC\IP^4$ gives us a family of metrics but of low dimension.
This problem is now solved.
Using our embedding by a basis of $N$ sections $s^I$, and
pulling back the Fubini-Study metric on $\IC\IP^{N-1}$, 
the embedding then leads to the K\"ahler potential
\be \label{eq:appendix_a1}
K = \log \sum_{I,\bJ} h_{I,\bJ} s^I \bs^\bj
\ee
where $s^I$ is a basis of $N=h_0(\CL)$ holomorphic sections.  
This gives us an $N^2$ real dimensional
family of metrics parameterized by the hermitian matrix $h_{I,\bJ}$.

\subsection{ Feed-forward networks }

Let us briefly review the definition of a feed-forward network (FFN, also called
MLP for multilayer perceptron).  It is a parameterized function
\be \label{eq:defF}
F_w : \CX \rightarrow \CY ,
\ee
with an input $x\in\CX \cong \IR^D$ and an output $y\in \CY\cong \IR^{D'}$ (we will generally take $D'=1$).
We can define it as the composition of a series of functions
\be \label{eq:defMLP3}
F_w = W^{(d)} \circ \theta \circ W^{(d-1)} \circ \ldots \circ
\theta \circ W^{(1)} \circ \theta \circ W^{(0)} ,
\ee
where the $W^{(i)}$'s are general linear transformations, and $\theta$
is a nonlinear function which acts independently on each vector component.
Explicitly,
\be
W^{(i)} : \IR^{D_i} \rightarrow \IR^{D_{i+1}} : v\rightarrow W^{(i)}\,v
\ee
where the $W^{(i)}$ on the right is a rectangular matrix,
$D_0=D, D_{d+1}=D'$ and the dimensions $D_1,D_2,\ldots$ of the intermediate
vector spaces can be freely chosen.  The function $\theta$ can be written
as a sum over a basis $e_a$ as
\be
\theta : \IR^{D_i} \rightarrow \IR^{D_{i}} : v\rightarrow 
\sum_{a=1}^{D_i} e_a \theta(v_a) 
\ee
where the $\theta(x)$ on the right, called the ``activation function,'' 
maps $\IR\rightarrow \IR$.
Two popular choices are
$\theta(x)=\tanh x$, and the ``ReLU'' function 
\be
\theta_{ReLU}(x) = \begin{cases} x, x\ge 0 \\ 0, x<0 \end{cases}.
\ee
One generally refers to a combination $\theta\odot W$ as a layer, with 
the final layer $W^{(d)}$ being an exception in not having $\theta$.
The term ``unit'' is sometimes used to denote the computation which
takes an input $v$ and produces a single component of $(\theta\circ W)(v)$,
so this network will have $D_1+\ldots+D_d$ units.
The number of layers $d+1$ is the depth.

It has been shown that feed-forward networks can approximate arbitrary functions,
including complex functions \cite{voigtlaender_universal_2020,kim2003approximation}. 
This is the case even for depth two ($d=1$) \citep{Cybenko1989}, but in this case one can need an exponentially
large number of units, as would be the case for simpler methods of interpolation (the ``curse of dimension'').
By using more layers, one can gain many advantages -- complicated functions can be represented with
many fewer units, and local optimization techniques are much more effective.  
How exactly this works is not
well understood theoretically and there are many interesting observations and hypotheses as to how these
advantages arise.

\subsection{ Supervised learning, sampling and data }
\label{appendix:learn}

In supervised learning, we have a data set of $N_{data}$ items, each of which is an input-output pair $(x_n,y_n)$.
These are supposed to be drawn from a probability distribution $\CP$ on $\CX\times\CY$.
The goal is to choose the function \equationref{eq:defF} from $\CX$ to $\CY$
which best describes the general relation $\CP$ between input and output,
in the sense that it minimizes some definition of the expected error (an objective or ``loss'' function).
The procedure of making this choice given the data set
is called training the network.

A simple choice of objective function is the mean squared error (MSE), 
\be \label{eq:mse}
\CE = \E{\CP}{ \left(f_w(x) - y\right)^2 }.
\ee
If we estimate this by evaluating it on our data set, we get the training error
\be \label{eq:train}
\CE_{train} =  \frac{1}{N_{data}} \sum_{n=1}^{N_{data}}{ \left(f_w(x_n) - y_n\right)^2 }.
\ee
Alternatively, one can also use the mean absolute percentage error (MAPE),
\be \label{eq:mape}
\CE = \E{\CP}{ \frac{\left|f_w(x) - y\right|}{y} }.
\ee

A standard ML training procedure is the following.  We start with an MLP as in \equationref{eq:defMLP},
with the weights initialized to random values -- in other words, we draw the $w$ from some distribution
independent of the data.  A common choice is for each matrix element $w_{i_m}^{(m),i_{m+1}}$ to be
an independent Gaussian random variable with mean zero and variance $1/\sqrt{D_m}$.  This choice is made so that
the expected eigenvalues of the weight matrix remain order one as we vary the $D_m$'s.

The next step is to minimize \equationref{eq:train} as a function of the weights.  A simple algorithm for this is
gradient descent, a stepwise process in which the weights at time $t+1$ are derived from those at $t$ as
\be \label{eq:gd}
w({t+1}) = w({t}) - \epsilon(t) \frac{ \partial \CE_{train} } { \partial w }\bigg|_{w=w(t)} .
\ee 
While this will only find a local minimum, it works better for these problems than one might have thought.
One trick for improving the quality of the result is
to make the step size $\epsilon(t)$ decrease with time, according to a ``learning schedule''
chosen empirically to get good results for the task at hand.

A variation on this procedure is ``stochastic gradient descent'' or SGD.  
This is much like \equationref{eq:gd}
except that instead of evaluating the training error $\CE_{train}$ on the full data set, one evaluates it on
a subset (or ``batch'') of the data set, 
with the batch varied from one step to the next so that their union covers the full data set.  This was
originally done for computational reasons but it also turns out to produce a noise term with beneficial 
properties, for example in helping to escape local minima.  
There are also many variations on SGD as well as other optimization algorithms, each with advantages
for certain applications.  We will describe our methods in \S \ref{ss:impl}.

Once the optimization is deemed to have converged, one judges the results by estimating \equationref{eq:mse}.
This estimate must be made by using an independent data set from that used in training as otherwise we are
rewarding our model for matching both signal and noise.\footnote{In classification problems, one often uses networks
with many more parameters than data points and which can completely fit the dataset, so that the minimum
of $\CE_{train}$ is zero!  In this case $\CE_{train}$ is clearly a poor estimate for $\CE$.}
However in most applications we do not have any direct access to $\CP$, rather we only have an empirical
data set.  Thus one starts by dividing the full data set into disjoint ``training'' and ``testing'' subsets, evaluates
\equationref{eq:train} on the training set for training, and then evaluates the sum of errors over the testing set to estimate $\CE$. 
The final model can be very accurate, surprisingly so when compared to expectations from standard
statistical theory.  Let us cite \citet{Zhang2017,Belkin2019} 
as two influential recent papers which developed this point.

While our problem is not one of supervised learning, it will be useful to phrase it in terms as similar
as possible, so that we can most easily use ML software.  The workflow of the supervised learning task involves defining
a set of data points $(x_n,y_n)$ which are independent of the weights, repeated evaluation of the network at each $x_n$
to get a prediction $f(x_n)$ for the corresponding $y_n$, and optimization of an objective function which is
a sum of terms which each depend on a single data point.
The network is normally defined by concatenating layers, such as multiplication by a weight matrix
(a fully connected layer), application of an activation function, and so on.  These layers are implemented
in associated software libraries, such as Keras for Tensorflow.  As we explain in \S \ref{ss:genimp},
while we will have to implement some new layers for our problem, otherwise our workflow is the same.

\section{ Nearly singular Calabi-Yau threefolds }

As discussed in the main text, a hypersurface $M$ defined as the solutions to $f=0$
will be a manifold only if $\p f=0$ everywhere on $M$.
There is much to say about the singular case, but our computations 
will be for
non-singular manifolds.  Still, a manifold which is nearly singular in the
sense we now describe will have small cycles and a Ricci flat
metric with corresponding large
ratio of length scales, which is the leading effect controlling the
accuracy of our numerical metrics.  Here we give a heuristic discussion of this
dependence.

The generic singularity of a hypersurface is an ordinary double point (ODP).  In a small neighborhood
of a $D=3$ ODP singularity it looks like
\be\label{eq:conifold}
z_1^2 + z_2^2 + z_3^2 + z_4^2 = \epsilon
\ee
and the singular limit is $\epsilon\rightarrow 0$.  This is usually called the conifold singularity in the
string theory literature.  A Ricci-flat metric on this noncompact manifold is known \citep{Candelas1990}
and it looks like the total space of $T^*S^3$, with the volume of the $S^3$ shrinking to zero as $\epsilon^{3/2}$
when $\epsilon\rightarrow 0$.\footnote{This can be seen by writing the real and imaginary parts of
\equationref{eq:conifold} separately.  For $\epsilon>0$ real, the $S^3$ is the submanifold $\Im z_i=0$.}
Thus the smallest length scale on this noncompact CY is $L \sim \epsilon^{1/2}$.
For $\epsilon\ne 0$, the solution of $\nabla f=0$ is all $z_i=0$, and the distance (in the Euclidean metric)
from this point to the closest solution of $f=0$ is also $\epsilon^{1/2}$.  

In the limit $\epsilon\rightarrow 0$, the metric becomes singular, with K\"ahler potential
\be
K \sim \left(z_1^2 + z_2^2 + z_3^2 + z_4^2\right)^{4/3}
\ee
near the singularity.  This K\"ahler potential is a $C^2$ function for which
one expects the Fourier coefficients to fall off as $k^{-4}$, which is
consistent with the numerical results in \citet{Headrick2013}.
The K\"ahler form $\omega$ and the CY volume
form $d\mu_\Omega$ (\equationref{eq:Omegavol}) should then be $C^0$, so one does
not expect other numerical problems besides the large ratio of scales.

To identify the region described by \equationref{eq:conifold} and $\epsilon$ in our
equations such as \equationref{eq:def_f1} and \equationref{eq:def_f2}, we use an idea from \citet{Blum1998}.
This is to first define the distance $d_Z(f,\Delta_Z)$ from a given $f$ to $\Delta_Z$,
the subset of $\Delta$ for which $f(Z)=\partial f(Z)=0$ for some $Z\in\IC\IP^4$.
We then minimize over $Z$,
\be
d(f,\Delta) = \min_{Z\in M} d(f,\Delta_Z) .
\ee
To define $d_Z(f,\Delta_Z)$, note that the subset of $\CM$ for which $f(Z)=0$ is a linear subspace,
call it $V_Z$.  Thus, we can use distance to $\Delta_Z$ in the Fubini-Study metric restricted to $V_Z$.\footnote{This definition of distance depends on the choice of FS metric.
In \citet{Blum1998} it was used to get expectations under a probability measure on
hypersurfaces which also depended on this choice.  But for our application, this is a
deficiency.  The symmetric choice we made gives reasonable results, but
it would be interesting to remove this dependence, perhaps by using the balanced FS metric or
best approximate Ricci flat metric adapted to $M$.
}
Let $||f||_H$ be the norm of $f$ in this metric;
a convenient way to express it is
\be
||f||_H^2 = \CN \int_{\IC^5} e^{-|Z|^2} |f(Z)|^2 ,
\ee
while the constraints $\partial_i h=0$ are best expressed in terms of a kernel
\be
\CK_n(Z, \bZ) = H_{I\bar J} Z^I  \bZ^\bJ = \frac{1}{n!} (\sum_i Z_i \bZ_\bi)^n .
\ee

Now, let $c_i\in V$ be the constraints $\partial_i Z^I$, with inner products 
\bea
L_{i\bj} &=& c_i\cdot H\cdot c_\bj \\
&=& \frac{\p}{ \p Z^i} \frac{\p}{ \p \bZ^\bj} \CK_n(Z, \bZ) \bigg|_{Z=Z_0} \\
&=& \delta_{i\bj} \CK_{n-1}(Z, \bZ) + \bZ_i Z_\bj \CK_{n-2}(Z, \bZ) .
\eea
Then the minimal distance from $f$ to $\Delta_{Z_0}$
is the norm in the subspace given by the projection $P=(L^{-1})^{i\bj} H\cdot c_i c_\bj$.
\be
\sin^2 \theta = \frac{ \langle f, P f \rangle_H }{ ||f||_H^2 } \\
= \frac{ (L^{-1})^{i\bj} \p_i f \p_\bj\bar f |_{Z=Z_0}  }{ ||f||_H^2 } .
\ee
The denominator is independent of $Z_0$, and the matrix $\delta_{i\bj} + (n-1) \bZ_i Z_\bj/|Z|^2$ 
is easy to invert. This will give combinations $|Z^i\p_i f|^2 (n-1)/n=(n-1)n|f(Z_0)|^2=0$ and 
finally
\be\label{eq:defdist}
\sin \theta \propto d \equiv \min_{Z_0\in M} \frac{ |\p_i f(Z_0)|_H }{ ||f||_H |Z_0|^{n-1}} .
\ee
For our purposes it suffices to take the right hand side of this equation
as our definition of distance, and to compute it in examples using numerical
minimization.

\begin{figure}
    \centering
    \includegraphics[scale=0.6]{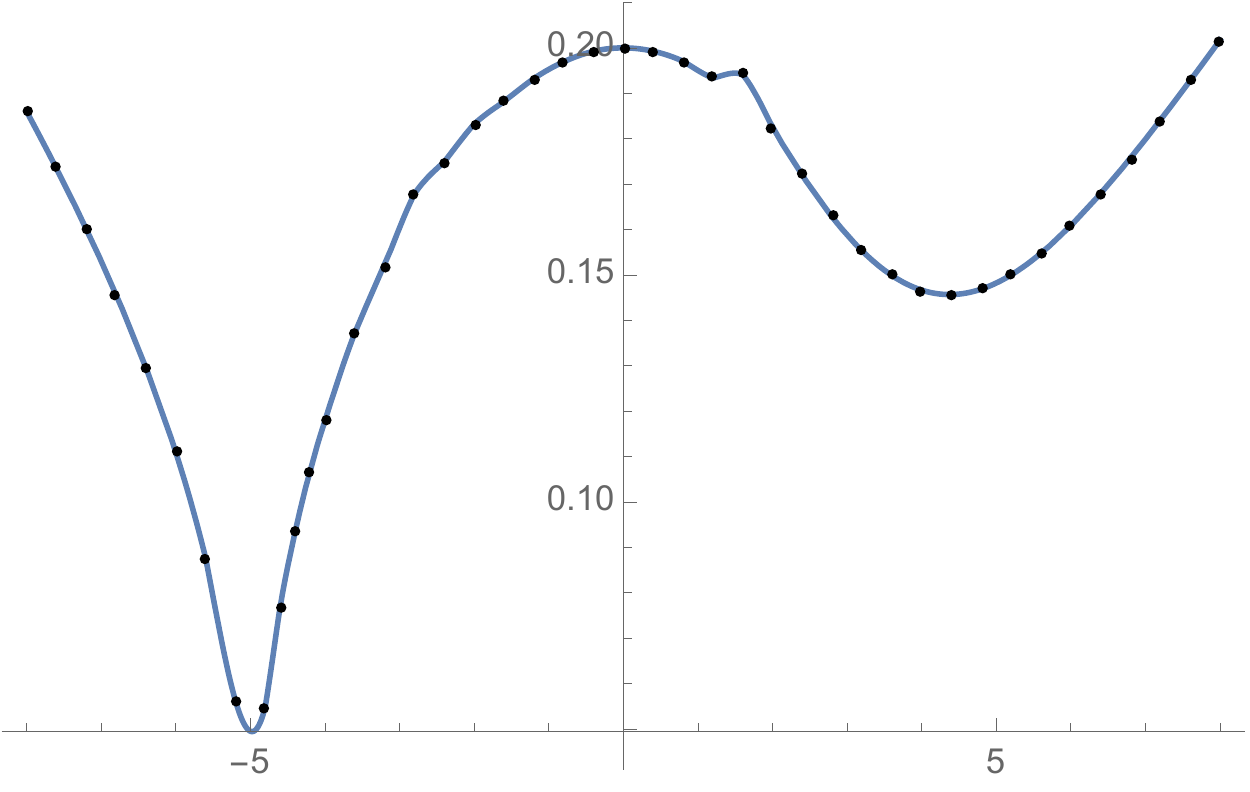}
    \caption{Distance to discriminant locus for the Dwork quintics,
    \eq{fermat}.  X axis is $\psi$, Y axis is $d$ in \eq{defdist}.}
    \label{fig:ddist}
\end{figure}

We plot \eq{defdist} for the Dwork quintics
in \figureref{fig:ddist}.  Besides the conifold point at $\psi=-5$,
there is a local minimum near $\psi=5$, which fits with the feature seen
in the plot of curvature versus $\psi$ in \citet{Cui2020}.  This is the point on the 
positive real axis closest to the conifold point, perhaps reached by following a
path like $\psi=5e^{i\theta}$.  It would be nice to
check this against the known
exact metric for this example \citet{Candelas1991}.  

Heat map plots of \eq{defdist} for the other defining functions appear in
\figureref{fig:dist1} and \figureref{fig:dist2}.

\end{document}